\documentclass[physrev,reprint,superscriptaddress,bibnotes]{revtex4-2}  

\usepackage[T1]{fontenc}
\usepackage[utf8]{inputenc} 
\usepackage[australian]{babel}

\usepackage[a4paper,centering,hmargin=1.75cm,vmargin=2cm]{geometry} 
\usepackage{amsmath,amssymb,graphicx,bm,microtype} 
\usepackage[colorlinks,allcolors=blue!50!black]{hyperref} 
\usepackage[all]{hypcap}
\usepackage{cleveref,braket,siunitx}
\usepackage[version=4]{mhchem}

\let\Re\relax \DeclareMathOperator{\Re}{Re}

\usepackage{tikz}
\newcommand{\blackdiamond}[1][fill=black]{\tikz [x=1.6ex,y=1.6ex,line width=0.1ex,line join=round] \draw  [#1]  (0.1,.5) -- (0.5,0.9) -- (0.9,.5) -- (0.5,0.1) -- (0.1,.5) -- cycle;}

\begin{document}

\title{Direct observation of geometric-phase interference in dynamics around a conical intersection}

\author{C.~H.~Valahu}
\thanks{These authors contributed equally to this work.}
\affiliation{School of Physics, University of Sydney, NSW 2006, Australia}
\affiliation{ARC Centre of Excellence for Engineered Quantum Systems, University of Sydney, NSW 2006, Australia}

\author{V.~C.~Olaya-Agudelo}
\thanks{These authors contributed equally to this work.}
\affiliation{ARC Centre of Excellence for Engineered Quantum Systems, University of Sydney, NSW 2006, Australia}
\affiliation{School of Chemistry, University of Sydney, NSW 2006, Australia}

\author{R.~J.~MacDonell}
\thanks{Current address: Department of Chemistry, Dalhousie University, Halifax, NS B3J 4H9, Canada}
\affiliation{ARC Centre of Excellence for Engineered Quantum Systems, University of Sydney, NSW 2006, Australia}
\affiliation{School of Chemistry, University of Sydney, NSW 2006, Australia}
\affiliation{University of Sydney Nano Institute, University of Sydney, NSW 2006, Australia}

\author{T.~Navickas}
\affiliation{School of Physics, University of Sydney, NSW 2006, Australia}
\affiliation{ARC Centre of Excellence for Engineered Quantum Systems, University of Sydney, NSW 2006, Australia}

\author{A.~D.~Rao}
\affiliation{School of Physics, University of Sydney, NSW 2006, Australia}
\affiliation{ARC Centre of Excellence for Engineered Quantum Systems, University of Sydney, NSW 2006, Australia}

\author{M.~J.~Millican}
\affiliation{School of Physics, University of Sydney, NSW 2006, Australia}
\affiliation{ARC Centre of Excellence for Engineered Quantum Systems, University of Sydney, NSW 2006, Australia}

\author{J.~B.~Pérez-Sánchez}
\affiliation{Department of Chemistry and Biochemistry, University of California San Diego, La Jolla CA, 92093, USA}

\author{J.~Yuen-Zhou}
\affiliation{Department of Chemistry and Biochemistry, University of California San Diego, La Jolla CA, 92093, USA}

\author{M.~J.~Biercuk}
\affiliation{School of Physics, University of Sydney, NSW 2006, Australia}
\affiliation{ARC Centre of Excellence for Engineered Quantum Systems, University of Sydney, NSW 2006, Australia}

\author{C.~Hempel}
\affiliation{School of Physics, University of Sydney, NSW 2006, Australia}
\affiliation{ARC Centre of Excellence for Engineered Quantum Systems, University of Sydney, NSW 2006, Australia}
\affiliation{Quantum Center, ETH Zurich, CH-8093 Zurich, Switzerland}
\affiliation{ETH Zurich-PSI Quantum Computing Hub, Paul Scherrer Institut, 5232 Villigen, Switzerland}

\author{T.~R.~Tan}
\email{tingrei.tan@sydney.edu.au}
\affiliation{School of Physics, University of Sydney, NSW 2006, Australia}
\affiliation{ARC Centre of Excellence for Engineered Quantum Systems, University of Sydney, NSW 2006, Australia}

\author{I.~Kassal}
\email{ivan.kassal@sydney.edu.au}
\affiliation{ARC Centre of Excellence for Engineered Quantum Systems, University of Sydney, NSW 2006, Australia}
\affiliation{School of Chemistry, University of Sydney, NSW 2006, Australia}
\affiliation{University of Sydney Nano Institute, University of Sydney, NSW 2006, Australia}

\begin{abstract}
Conical intersections are ubiquitous in chemistry and physics, often governing processes such as light harvesting, vision, photocatalysis, and chemical reactivity. They act as funnels between electronic states of molecules, allowing rapid and efficient relaxation during chemical dynamics. In addition, when a reaction path encircles a conical intersection, the molecular wavefunction experiences a geometric phase, which can affect the outcome of the reaction through quantum-mechanical interference. Past experiments have measured indirect signatures of geometric phases in scattering patterns and spectroscopic observables, but there has been no direct observation of the underlying wavepacket interference. Here, we experimentally observe geometric-phase interference in the dynamics of a wavepacket travelling around an engineered conical intersection in a programmable trapped-ion quantum simulator. To achieve this, we develop a technique to reconstruct the two-dimensional wavepacket densities of a trapped ion. Experiments agree with the theoretical model, demonstrating the ability of analog quantum simulators---such as those realised using trapped ions---to accurately describe nuclear quantum effects. 
\end{abstract}

\maketitle

Light drives molecular processes as important as photosynthesis, photocatalysis, and vision. Absorbing a photon promotes a molecule to an excited electronic state, triggering chemical dynamics and reactivity. The molecule will eventually return to the ground state; often, this relaxation happens on ultrafast (fs--ps) timescales at molecular geometries where two electronic energy surfaces have the same energy, known as conical intersections~\cite{Yarkony, Domcke, LarsonBook}. By acting as funnels between electronic states for the molecular wavefunction, conical intersections enable rapid non-radiative electronic transitions and have a decisive role in chemical dynamics, from charge-transfer processes to photochemical reactions~\cite{Domcke2}.

The path taken during molecular dynamics involving conical intersections can profoundly alter chemical reaction outcomes. In particular, a geometric phase~\cite{Berry} causes quantum interference of wavepackets encircling a conical intersection~\cite{Longuet, Mead-Truhlar, Schon, Ryabinkin}. Accounting for geometric phase is necessary in quantum chemistry calculations because the resulting interference changes the ratio of reactive and non-reactive outcomes in scattering cross-sections~\cite{Mead1980, Lepetit1990, Althorpe2006, Althorpe2008} and alters vibrational spectra~\cite{Kendrick1997, Applegate, Englman}. Indeed, recent experiments have detected indirect signatures of geometric phase in reactive scattering~\cite{Daofu,DaofuII}. An elegant proposal for revealing spectroscopic signatures of geometric phase involves interference signals from pairs of excitation pulses~\cite{Cina1990,Cina1991,Cina1993}, but it remains unimplemented due to challenging state preparation.

Conical intersections and the associated geometric phase are general phenomena that also appear in other branches of physics~\cite{LarsonBook}. In general, a conical intersection can form in any parameter-dependent quantum system where two energy surfaces cross. In molecules, the parameters are usually the normal modes of nuclear motion, but, in condensed-phase systems, conical intersections commonly arise as Dirac cones in reciprocal (momentum) space~\cite{LarsonBook}. These include the Dirac cones in graphene~\cite{Castro}, in superconductors~\cite{Ran_Superconductor}, and in the Rashba~\cite{Rashba} and Dresselhaus~\cite{Dresselhaus} treatments of spin-orbit coupling. 

An unambiguous observation of geometric-phase interference in wavepacket dynamics around a conical intersection remains an outstanding challenge. In a molecular or solid-state system, it would require a full reconstruction of the wavepacket dynamics on ultrafast timescales, which is possible in small molecules~\cite{Cina2008}, but has never been used to characterise geometric phase. 

Analog quantum simulators present a new opportunity to access quantum dynamics on laboratory-accessible timescales~\cite{Buluta, Roos, Aspuru, McArdle,Haeffner2018}. In such systems, a one-to-one correspondence between the degrees of freedom of the chemical or physical system and those of the simulator makes it possible to replicate the target dynamics in a controllable and measurable manner, as well as explore new parameter regimes in a controllable fashion.

Several controllable quantum systems have been proposed to engineer conical intersections and study signatures of geometric phase. 
Most of these quantum simulations have been performed in reciprocal space to simulate solid-state systems, including geometric phases around Dirac points~\cite{Bloch2015, Brown2022, LarsonBook}.
Theoretical proposals for simulating molecular conical intersections have included using trapped Rydberg ions to simulate electronic populations~\cite{Gambetta}, circuit quantum electrodynamics to simulate emission spectra~\cite{Dereli}, and cavity quantum electrodynamics to simulate collapse-revival characteristics of a spreading wavepacket~\cite{Larson}. To date, the only experimental quantum simulation of a chemical conical intersection demonstrated branching between different photochemical reaction products with strong dissipation~\cite{Christopher2022}.

\begin{figure*}[t]
	\centering
	\includegraphics[width=\textwidth]{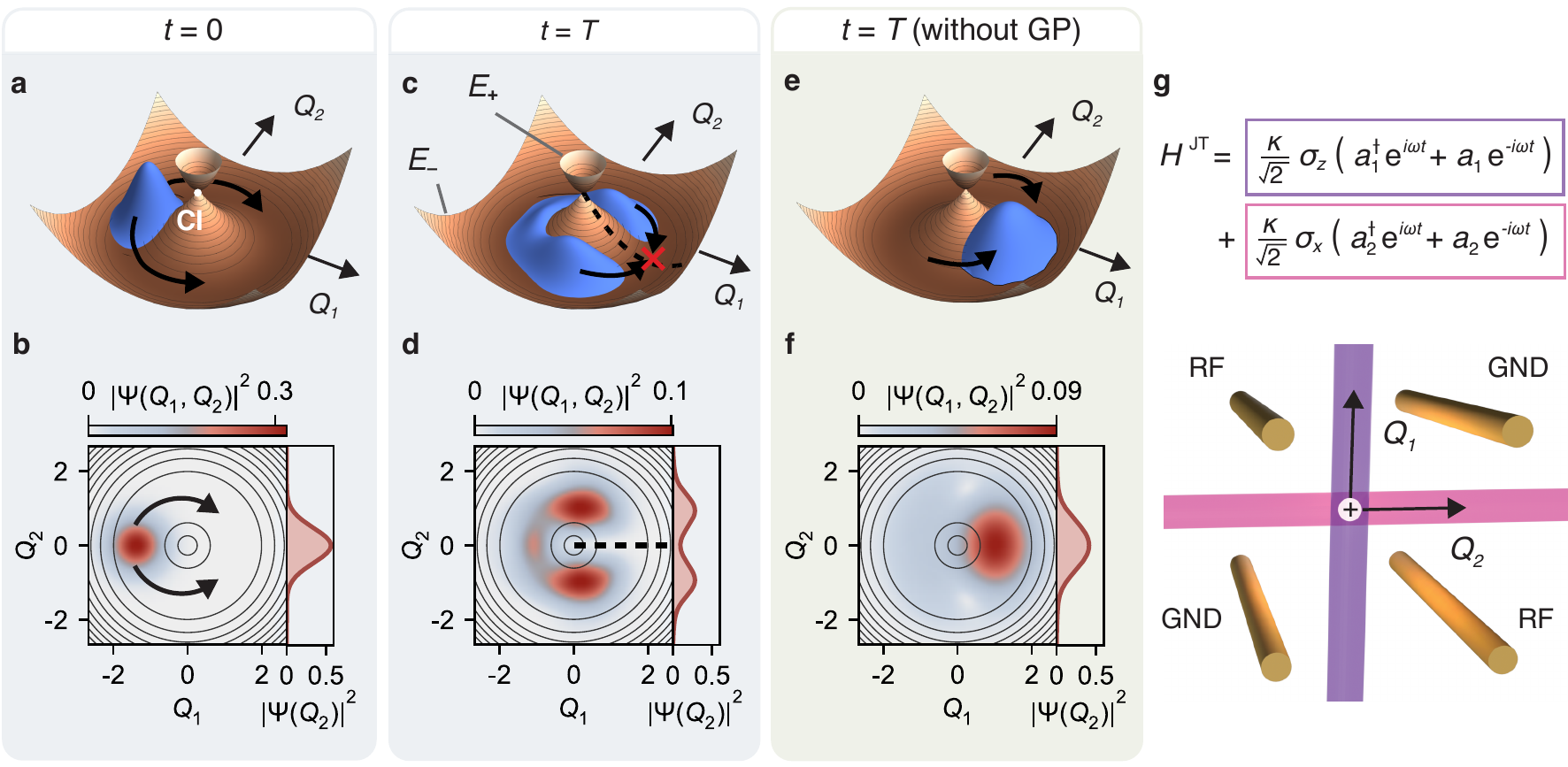}
	\caption{\textbf{Directly detecting a geometric phase through wavepacket interference.}
	\textbf{a},~A motional wavepacket is initially displaced to the minimum of the potential energy surface, after which it begins to encircle the conical intersection, denoted CI.
	\textbf{b},~Initial wavepacket density in 2D (left), and integrated over $Q_1$ (right).
	\textbf{c},~After sufficient time evolution, the two components of the wavepacket destructively interfere due to geometric phase, giving a nodal line along $Q_2=0$ (dotted line). 
	\textbf{d},~Motional wavepacket density at the maximum interference time $T$.
	\textbf{e},~If the geometric phase were neglected, the two wavepacket components would interfere constructively.
	\textbf{f},~Density at $t=T$ with geometric phase neglected.
	Contours in b, d, and f correspond to the potential energy surface $E_-$.
	\textbf{g},~The Jahn-Teller Hamiltonian $H^\mathrm{JT}$ is engineered in an ion-trap quantum simulator with a single \ce{^171Yb+} ion. The ion (white sphere) is confined in a Paul trap and $H^\mathrm{JT}$ is realised using two simultaneous laser-induced interactions (purple and pink, corresponding to colour-coded terms in $H^\mathrm{JT}$).
	}
	\label{fig:GPExp}
\end{figure*}

Here, we present the observation of the destructive interference caused by geometric phase during dynamics of a wavepacket around a conical intersection. We implement a controllable conical intersection by engineering a Jahn-Teller Hamiltonian in a trapped-ion quantum simulator that employs a mixed-qudit-boson (MQB) encoding in which both the ion's electronic and motional degrees of freedom are used~\cite{MacDonell2021}. This work is not merely a simulation of geometric phase: the ion is a real, observable, and measurable quantum system that undergoes conical-intersection dynamics, allowing us to directly observe the geometric-phase interference of its motional wavepacket. To this end, our experiment introduces a resource-efficient reconstruction method to image the wavepacket's probability density, directly showing the destructive self-interference as the wavepacket encircles a conical intersection. Experimental measurements match theoretical predictions, demonstrating the utility of quantum simulators to give insights into properties that have otherwise been impossible to measure directly for chemical systems. 

\section*{Results}

In an MQB simulator~\cite{MacDonell2021}, the electronic and vibrational degrees of freedom that are to be simulated are represented in a qudit and a set of bosonic modes. We realise a conical intersection using an \ce{^171Yb+} ion confined in a Paul trap, where two vibrations are encoded directly in the ion's transverse vibrational modes (B$_1$ and B$_2$), while two electronic states are encoded in the ion's qubit (qudit with $d=2$) states comprising the two hyperfine levels of the \ce{^2S_{1/2}} ground state (detailed in Methods). This approach has recently been employed to predict molecular spectra using time-domain simulations~\cite{MacDonell2022}, and provides resource-scaling advantages relative to conventional methods of quantum simulation~\cite{MacDonell2021}.

\begin{figure*}[t]
    \centering
   \includegraphics[width=1\textwidth]{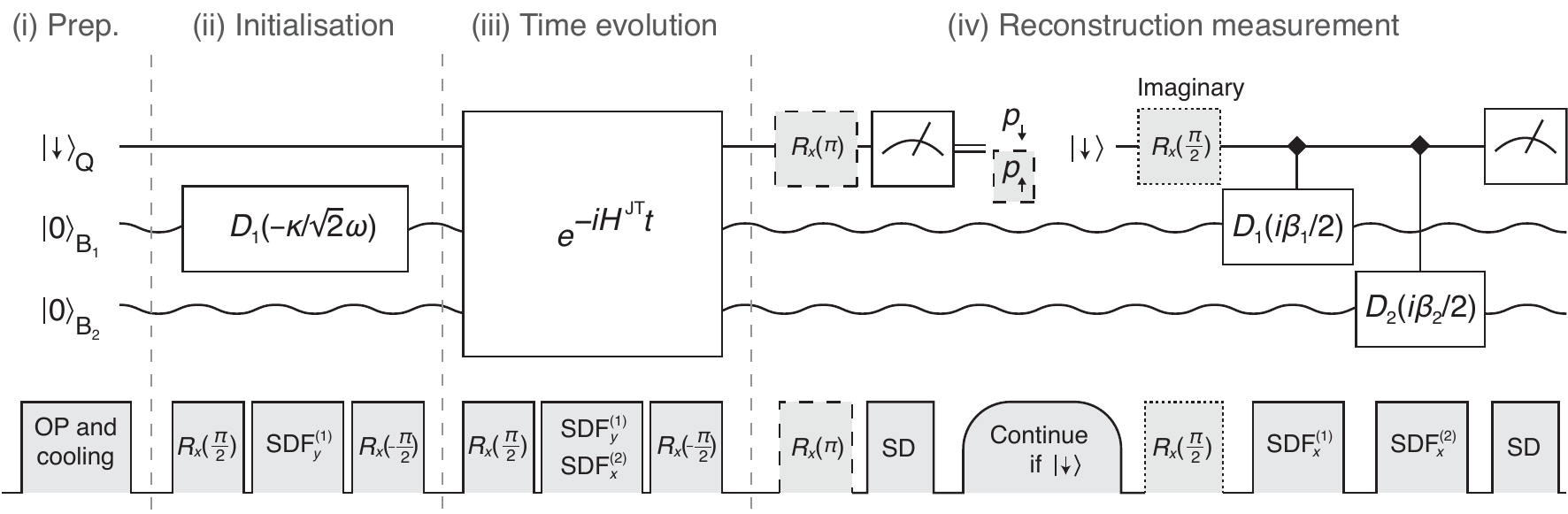}
    \caption{\textbf{Experimental protocol for geometric-phase dynamics simulation and wavepacket reconstruction.}
    \textbf{(top)}~Quantum circuit diagram for a single-trapped-ion simulator, consisting of a qubit (Q, solid line) and 2 bosonic modes (B$_1$, B$_2$, wavy lines), and \textbf{(bottom)}~corresponding experimental pulse sequence. 
    \textbf{(i)}~Preparation of fiducial states by optical pumping and cooling. 
    \textbf{(ii)}~Initialisation: B$_1$ is displaced by $D_1(-\kappa/\sqrt{2}\omega)$, implemented using an SDF pulse with surrounding qubit $\pi/2$ pulses to map to the correct basis.
    \textbf{(iii)}~Time evolution under the Jahn-Teller Hamiltonian for duration $t$, implemented using two simultaneous SDF pulses in different bases, acting on the two bosonic modes.
    \textbf{(iv)}~Reconstruction measurement: a state detection (SD) pulse collapses the qubit state, and the circuit proceeds only if the measured state was $\ket{\downarrow}$ (rounded shape). An additional single-qubit pulse (dashed) is introduced to retrieve information entangled with the $\ket{\uparrow}$ state. The qubit probabilities $p_\downarrow$ and $p_\uparrow$ are calculated from the mid-circuit measurement outcomes. Controlled displacements acting in the $\sigma_x$ basis (denoted~$\blackdiamond$) map the motional degrees of freedom onto the qubit, allowing the real part of the characteristic function to be measured. Its imaginary part is obtained using an additional $R_x(\pi/2)$ pulse prior to the controlled displacements (dotted).
    Successive $R_x$ rotations shown separately in the circuit are combined into a single pulse in the experiment.
    }
    \label{fig:experimental_pulse_sequence}
\end{figure*}

To demonstrate geometric-phase interference, we implement the $E\otimes e$ Jahn-Teller model~\cite{Longuet}, a standard model of geometric-phase effects in molecules~\cite{Bersuker, Schon}. It consists of two electronic states coupled with two vibrational modes, described by the potential energy
\begin{equation}
    V^\mathrm{JT} = \frac{\omega}{2} (Q_1^2 + Q_2^2)
    + \kappa (\sigma_z Q_1 + \sigma_x Q_2),
\end{equation}
where $\sigma_x$ and $\sigma_z$ are the Pauli matrices acting on the electronic states and $Q_j = (a_{j}^{\dag} + a_{j})/\sqrt{2}$ is the dimensionless position coordinate for the $j$th vibrational mode, with creation and annihilation operators $a_{j}^{\dag}$ and $a_{j}$.
$\kappa$ is the vibronic coupling strength, and $\omega$ is the frequency of both vibrational modes. The Jahn-Teller Hamiltonian is given by $H^\mathrm{JT} = \omega (P_1^2 + P_2^2)/2 + V^{JT}$, where $P_j$ is the conjugate momentum of $Q_j$.
We set $\hbar=1$ throughout.

Diagonalisation of $V^\mathrm{JT}$ in the electronic basis leads to cylindrically symmetric potential energy surfaces along $Q_1$ and $Q_2$, with energies $E_\pm = \omega (Q_1^2 + Q_2^2)/2 \pm \kappa \sqrt{Q_1^2 + Q_2^2}$ (see \cref{fig:GPExp}).
The conical intersection is present at the point of highest symmetry ($Q_1 = Q_2 = 0$), where the two potential energy surfaces are degenerate.
The minimum of $E_-$ occurs where $Q_1^2 + Q_2^2 = (\kappa / \omega)^2$.

The effects of geometric phase on dynamics around a conical intersection can be directly observed from the motional probability density, \cref{fig:GPExp}a--d. As the initial wavepacket, we choose the ground state of the non-interacting vibrational Hamiltonian, $H_0=\omega(a^\dagger_1 a_1+a^\dagger_2 a_2)$, displaced to the potential-energy minimum at $Q_1 = -\kappa/\omega$, $Q_2=0$ (\cref{fig:GPExp}a--b). During the time evolution, the wavepacket splits into two components evolving in opposite directions around the conical intersection. The two components overlap at $Q_1 > 0$, causing destructive interference at the nodal line $Q_2=0$, where their equal and opposite geometric phases lead to a vanishing density (\cref{fig:GPExp}c--d). By contrast, if geometric phase were disregarded, the two wavepacket fragments would interfere constructively, reaching maximum amplitude at $Q_2=0$ (\cref{fig:GPExp}e--f).

To map the Jahn-Teller model onto the MQB simulator, we rewrite $H^\textrm{JT}$ in the interaction picture with respect to $H_0$,
\begin{align}
    H_\mathrm{I}^\mathrm{JT} = {} & \frac{\kappa}{\sqrt{2}} \sigma_z (a_{1}^{\dag} e^{i\omega t} + a_{1} e^{-i\omega t}) + \nonumber \\ 
    & \frac{\kappa}{\sqrt{2}} \sigma_x (a_{2}^{\dag}e^{i\omega t} + a_{2} e^{-i \omega t}), \label{eq:mol_int}
\end{align}
which can be implemented using tunable light-atom interactions to enact qubit-boson couplings. 
We achieve this implementation using a coherent state-dependent force (SDF) enacted by stimulated Raman transitions driven with a \SI{355}{nm} pulsed laser~\cite{Monroe1996, Mizrahi2013}. Driving transitions near bosonic mode $j$ leads to the Hamiltonian
\begin{equation}
    H^\mathrm{SDF}_{j,\phi_s}(\delta,\Omega, \phi_m) = \frac{\Omega}{2} \sigma_{\phi_s} ( a_j^\dagger e^{i(\phi_m + \delta t )} + a_j e^{-i(\phi_m + \delta t)} ),
    \label{eq:H_sdf_main}
\end{equation}
where $\sigma_{\phi_s} = \sigma_x\cos\phi_s + \sigma_y\sin\phi_s$ and $\phi_s$ and $\phi_m$ are the phases associated with the qubit and the bosonic mode, respectively (see Methods). $\Omega$ and $\delta$ are the Rabi frequency and detuning of the laser from the bosonic mode, respectively. We use the notation $H^\mathrm{SDF}_{j,x}$ and $H^\mathrm{SDF}_{j,y}$ for SDF interactions where $\phi_s = 0$ and $\phi_s=\pi/2$, respectively. Interactions in the $\sigma_z$ basis are obtained using a qubit basis rotation, $H^\mathrm{SDF}_{j,z} = R_x(\pi/2) H^\mathrm{SDF}_{j,y} R_x(-\pi/2)$, where $R_x(\theta)$ are driven qubit rotations around the Bloch sphere. 
$H^\mathrm{JT}$ can then be implemented in a programmable way using two simultaneous SDFs (see \cref{fig:GPExp}g),
\begin{equation}
    H^\mathrm{JT}_\mathrm{I} = H^\mathrm{SDF}_{1,z}(\omega, \sqrt{2}\kappa, 0) + H^\mathrm{SDF}_{2, x} (\omega, \sqrt{2}\kappa, 0).
    \label{eq:JT_SDF}
\end{equation}
The parameters $\kappa$ and $\omega$ are chosen to produce a clear wavepacket interference. To achieve this, $\kappa/\omega$ should be large enough that the wavepacket prepared at the minimum of the potential energy surface (at $Q_1 = -\kappa/\omega$) has negligible overlap with the conical intersection. However, $\kappa/\omega$ should also be kept small enough to mitigate vibrational decoherence that increases with larger vibrational excitations. To balance these considerations, we choose $\kappa/\omega=1.5$, for which the wavepacket has only 1.7\% of the density at $Q_1 \geq 0$. $\kappa$, implemented by adjusting the Rabi frequency, is maximised to increase the speed of the dynamics; its value is constrained by the available SDF-laser power to $\kappa = 2\pi \times \SI{1.00}{kHz}$, yielding $\omega=2\pi\times\SI{667}{Hz}$. With these parameters, the wavepackets are expected to experience the greatest geometric-phase interference at $T=\SI{1.59}{ms}$, which was computationally predicted as half the time at which the width of the probability density is minimized.

\begin{figure*}[t]
    \centering
   \includegraphics[width=1\textwidth]{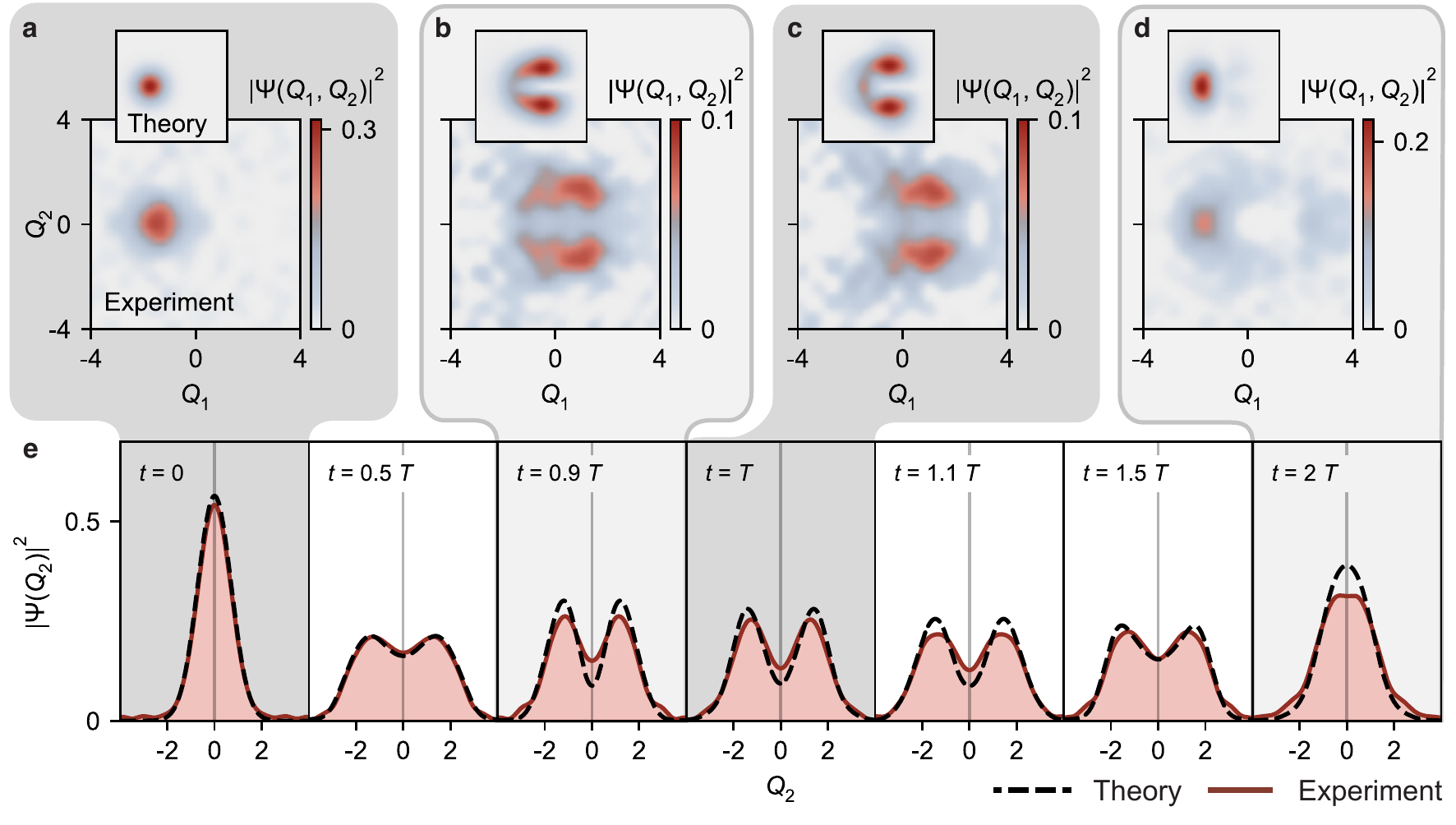}
    \caption{\textbf{Wavepacket dynamics around an engineered conical intersection.}
    \textbf{a-d}~Reconstructed two-dimensional motional densities of the \ce{^171Yb+} ion at times $t=0$, $0.9T$, $T$, and $2T$, with theoretical predictions in the insets. The nodal line at $Q_2=0$, $Q_1>0$ is a direct signature of geometric-phase interference.
    \textbf{e}~One-dimensional motional densities (with the $Q_1$ coordinate integrated out) at more values of $t$. Geometric-phase interference causes the dip at $Q_2=0$ around $t=T$.
    The motional densities are obtained using \cref{CharacteristictoProbDensity} (and its one-dimensional equivalent) from the characteristic functions, $\chi(i\beta_1, i\beta_2)$ and $\chi(i\beta_2)$, measured using the circuit in \cref{fig:experimental_pulse_sequence} (see Methods for details).
    }
    \label{fig:density_movie}
\end{figure*}

We probe the dynamics of the geometric phase around the conical intersection by reconstructing the ion's motional probability densities at different evolution times $t$. The experimental sequence consists of four stages, shown in \cref{fig:experimental_pulse_sequence}. (i)~Preparation of the qubit and cooling of the vibrational modes to their ground states is achieved by optical pumping, Doppler cooling, and sideband cooling. 
(ii)~Initialisation consists of displacing B$_1$ to $Q_1 = -\kappa/\omega$ by applying an SDF interaction $H^\mathrm{SDF}_{1, z}(0,\Omega_1, \pi/2)$ for a duration $\tau$. This applies the displacement operator $D_1(-\Omega_1\tau/2)$, where $\Omega_1$ and $\tau$ are chosen to implement $D_1(-\kappa/\sqrt{2}\omega)$.  
(iii)~Evolution of the system under $H^\mathrm{JT}_I$ is achieved by applying the two simultaneous SDF interactions of \cref{eq:JT_SDF} for an experimentally variable duration $t$. 
(iv)~Reconstruction of the joint densities of B$_1$ and B$_2$ is achieved by measuring the characteristic function
\begin{equation}
    \chi(i\beta_1, i\beta_2) = \bra{\Psi}D_1(i\beta_1)D_2(i\beta_2) \ket{\Psi},\label{CharacteristictoProbDensity}
\end{equation}
where $\ket{\Psi}$ is the total wavefunction of the system, and $\beta_1$ and $\beta_2$ are real numbers. See Methods for details.

The joint probability densities are reconstructed using the circuit in \cref{fig:experimental_pulse_sequence}.  Two SDF pulses are sequentially applied on B$_1$ and B$_2$, and $\chi(i\beta_1, i\beta_2)$ is scanned over $\beta_1$, $\beta_2$. These measurements yield the joint probability density via the Fourier transform of the measured characteristic function
\begin{equation}
    |\Psi(Q_1, Q_2)|^2 = \! \iint \frac{d \beta_1 d \beta_2}{2\pi^2} e^{- i \sqrt{2} (Q_1 \beta_1 + Q_2 \beta_2)} \chi(i\beta_1, i\beta_2).
    \label{eq:2d_density_function}
\end{equation}

In further detail, we measure $\chi(i\beta_1, i \beta_2)$ by mapping information from the multimode bosonic system onto the qubit using SDF pulses, moving beyond previous works on direct single-mode~\cite{Leibfried1996, Gerritsma2010, Johnson2015, Home2020} and indirect multimode reconstructions~\cite{Jia2022}. The reconstruction consists of preparing the qubit in $\ket{\downarrow}$ and applying two successive SDF interactions, $H^\mathrm{SDF}_{1,x}(0, \Omega_2, 0)$ and $H^\mathrm{SDF}_{2,x}(0,\Omega_2, 0)$ with durations $\tau_1$ and $\tau_2$. Doing so results in controlled displacements $D_1(i\beta_1/2)$ and $D_2(i\beta_2/2)$, where $\beta_j = \Omega_2 \tau_j$. $\chi(i\beta_1, i\beta_2)$ is measured for different values of $\beta_1$ and $\beta_2$ by varying $\tau_1$ and $\tau_2$. We reconstruct the characteristic functions of the bosonic modes entangled with the $\ket{\downarrow}$ and $\ket{\uparrow}$ qubit states independently. Reconstructing the $\ket{\downarrow}$ component is achieved by adding a mid-circuit measurement which projects out the $\ket{\uparrow}$ component (see Methods). The experiment is repeated with an additional $R_x(\pi)$ pulse prior to the mid-circuit measurement to reconstruct the $\ket{\uparrow}$ component. The qubit probabilities $p_\downarrow$ and $p_\uparrow$ are calculated from the success rate of the mid-circuit measurement. After the displacements, measuring the qubit in the $\sigma_z$ basis gives the real part of the characteristic function, $\langle \sigma_z \rangle = \Re\chi(i\beta_1, i\beta_2)$. Repeating the experiment with an additional $R_x(\pi/2)$ pulse prior to the displacements gives the imaginary part, after which the full $\chi(i\beta_1, i \beta_2)$ is obtained by adding the real and imaginary parts associated with both $\ket{\downarrow}$ and $\ket{\uparrow}$.  

The reconstructed probability densities in \cref{fig:density_movie}a--d demonstrate a direct measurement of the wavepacket interference caused by the geometric phase. At $t=0$, the initial wavepacket is prepared at $(Q_1, Q_2) = (-1.5, 0)$. As the wavepacket evolves around the conical intersection, the nodal line becomes visible at $Q_2=0$ and is most pronounced at $t=T$; this is a direct observation of destructive interference due to geometric phase. Finally, at $t=2T$, the two wavepackets recombine close to their initial position. The experimental results agree well with theoretical predictions, reproducing key features of interference and wavepacket recombination.

Further quantitative insight may be gained from the 1-dimensional density $|\Psi(Q_2)|^2$, obtained by omitting the $D_1(i\beta_1)$ displacements from the reconstruction procedure discussed earlier. In this case, the measurements scanned over $\beta_2$ are Fourier-transformed to give $|\Psi(Q_2)|^2 = \int d \beta_2 \,e^{- i \sqrt{2} Q_2\beta_2} \chi(i\beta_2)/\sqrt{2}\pi$. In \cref{fig:density_movie}e, we present $|\Psi(Q_2)|^2$ for seven different evolution times. A comparison of experiment and theory shows excellent agreement in the shape and amplitude of the measured density function. We attribute minor discrepancies to the dephasing of the bosonic modes, miscalibrations such as uncompensated AC Stark shifts, and technical imperfections in the protocol implementation.

\section*{Discussion}

Our approach avoids the limitations of direct experiments on molecular systems, where only few observables---such as spectra and scattering cross sections---can be measured. 
Instead, a fully controllable quantum device---such as an ion-trap MQB simulator~\cite{MacDonell2021}---can, in principle, read out any observable; as we showed here, this includes the full two-dimensional density of the \ce{^171Yb+} ion as it moves in space and time. A further advantage comes from the ratio ($r$) of the ion's natural timescale (ms) and the measurement speed (ns), leading to an increase in the observable timing resolution of $r\sim10^6$. This improves the achievable resolution of chemical-dynamics measurements relative to ultrafast observations.

A key general feature of quantum simulations is their programmability~\cite{Hayes_2014}. Our work is a simulation of the dynamics of the Jahn-Teller model, which is often used to describe molecular systems. In an MQB simulator, the qudit-boson interaction is controllable, meaning that the same device can be programmed to simulate different molecular systems, solid-state systems, or theoretical models that do not occur naturally.
In particular, our geometric-phase simulator could be used to simulate dynamics in molecules with conical intersections where the interactions are not as symmetric as in $H^\mathrm{JT}$, such as the general quadratic vibronic-coupling Hamiltonian~\cite{MacDonell2021}.

Like any analog simulation---quantum or classical---our approach is ultimately limited by noise and uncorrected errors. In our experiment, the main sources of decoherence and dissipation are motional dephasing and motional heating~\cite{Brownnutt2015, MacDonell2022} (see Methods). However, in MQB simulations of molecular processes, noise can be characterised and even amplified in order to create a realistic model of molecular environments, such as collisions in solution. Since our Jahn-Teller experiment shows only weak effects of decoherence over the full period $2T$, we would need to inject additional noise to simulate conical-intersection dynamics of real molecules in chemically realistic situations (i.e., other than a single molecule in vacuum). In scaling up to larger molecules, the ability to simulate dissipation would allow us to probe regimes in nonadiabatic dynamics that are among the most difficult to simulate on conventional computers~\cite{MacDonell2021}.

Our methodology for probability density reconstruction enables scalability and resource efficiency. Early techniques for motional-state tomography were performed in the Fock basis~\cite{Leibfried1996, Kienzler2016, Jia2022}, a process that requires many measurements if full motional densities are sought. 
More recently, wavepacket-reconstruction methods were developed based on the direct measurement of the characteristic function, significantly reducing the number of necessary measurements~\cite{Home2020}.
Our approach builds on the latter techniques, but has two additional advantages. 
First, we extend the characteristic-function method to multimode probability density reconstruction, while retaining both the requirement of few measurements and the ability to use one readout qubit.
Second, using a mid-circuit measurement allows us to reuse the simulation qubit for the reconstruction, without any ancilla qubits. 

We have recently become aware of related simultaneous work on simulating a conical intersection using a chain of trapped ions~\cite{Whitlow2022}. The system was adiabatically driven to its vibronic ground state, whose reconstructed two-dimensional density showed a node attributed to geometric phase. This work is complementary to ours in several ways: it focused on signatures of geometric phase in the ground state, not in the dynamics; it used Trotterised time evolution, while we drove the two interactions simultaneously; and it used an ancilla qubit in the reconstruction, while we used the mid-circuit-measurement approach discussed earlier. 

In conclusion, our experiment represents the direct observation of wavepacket interference caused by geometric phase in dynamics around a conical intersection. Our approach to quantum simulation using an MQB trapped-ion system makes chemical dynamics that are otherwise unmeasurable directly accessible in the laboratory. This is a key demonstration of the utility of small-scale quantum computational devices to offer practical insights into chemical dynamics and resolve intractable problems in chemical physics.

\begin{acknowledgments}
We thank Jacob Whitlow and Kenneth Brown for valuable discussions. 
We were supported by the U.S. Office of Naval Research Global (N62909-20-1-2047), by the U.S. Army Research Office Laboratory for Physical Sciences (W911NF-21-1-0003), by the U.S. Intelligence Advanced Research Projects Activity (W911NF-16-1-0070), by Lockheed Martin, by the Australian Government's Defence Science and Technology Group, by the Sydney Quantum Academy (VCO, ADR, MJM, and TRT), by a University of Sydney-University of California San Diego Partnership Collaboration Award (JBPS, JYZ, and IK), by H.\ and A.\ Harley, and by computational resources from the Australian Government's National Computational Infrastructure (Gadi) through the National Computational Merit Allocation Scheme.
\end{acknowledgments}

\section*{Author Contributions Statement}

RJM, IK, CH and TRT conceived the original idea; VCO, RJM, JBP, JY, and IK developed the theoretical methods; CHV, TN, ADR, MJM, and TRT developed and carried out the experiments; CHV, VCO, TRT, and IK wrote the manuscript with feedback from all authors; all authors discussed the results and interpreted the data.

\section*{Competing Interests Statement}

The authors declare no competing interests.

\section*{Methods}

\setcounter{figure}{0}
\renewcommand{\figurename}{Extended Data Fig.}

\subsection*{Experimental setup}

The \ce{^{171}Yb^+} ion is confined in a Paul trap with radial mode oscillation frequencies of $2\pi \times \SI{1.34}{MHz}$ and $2\pi \times \SI{1.47}{MHz}$, corresponding to bosonic mode B$_1$ and B$_2$. The qubit is encoded in the two magnetically insensitive hyperfine levels of the \ce{^2S_{1/2}} ground state, where we assign the labels $\ket{\downarrow} = \ket{F=0, m_F=0}$ and $\ket{\uparrow} = \ket{F=1, m_F = 0}$.  

We use two laser beams derived from a \SI{355}{nm} pulsed laser to coherently control the qubit and bosonic modes via stimulated Raman transitions within the \ce{^2S_{1/2}} ground state. The two Raman beams are orthogonal to one another, and configured so that they can be coupled to both radial vibrational modes. Each Raman beam passes through an acousto-optical modulator (AOM), which allows the phase, frequency and amplitude of the beam to be adjusted by altering the RF signal driving the AOM. One of the RF signals is generated by an arbitrary waveform generator (Keysight M8190A), allowing multiple phase-coherent tones to be imprinted on one of the laser beams. We ensure phase coherence between all pulses in the experimental sequence by tracking the phase (relative to the beginning of the pulse sequence) and applying appropriate corrections.

By tuning the frequency difference of the Raman beams, one can drive carrier, red- and blue-sideband transitions. Qubit rotations are obtained by driving carrier transitions, while an SDF $H^\mathrm{SDF}_{j,\phi_s}(0, \Omega, \phi_m)$ arises from combining the red- and blue-sideband transitions. Applying this interaction for a duration $\tau$ with the qubit in an eigenstate of $\sigma_{\phi_s}$ displaces bosonic mode $j$ by $D_j(\alpha) = \exp(\alpha a^\dagger_j - \alpha^* a_j)$, where $\alpha = - i\Omega \tau e^{i\phi_m}/2$. The amplitude and phase-space direction of the displacement are adjusted by varying $\tau$ and $\phi_m$, respectively. 

\subsection*{Experimental protocol}

\textit{Preparation.} The bosonic modes are cooled in two stages. First, they are Doppler cooled using a \SI{369.5}{nm} laser red-detuned from the \ce{^2S_{1/2}} $\rightarrow$ \ce{^2P_{1/2}} transition.
Second, resolved sideband cooling is used to reach their motional ground states, achieving temperatures of $\bar{n} = 0.04$ measured via sideband thermometry \cite{Monroe1995}. The qubit is prepared in its ground state $\ket{\downarrow}$ via optical pumping, using another \SI{369.5}{nm} laser resonant with the \ce{^2S_{1/2}} $\ket{F=1}$ $\rightarrow$ \ce{^2P_{1/2}} $\ket{F=1}$ transition. 

\textit{Initialisation.} To initialise B$_1$, we apply an SDF interaction $H^\mathrm{SDF}_{1,y}(0, \Omega_1, \pi/2)$ for a duration $\tau$, which gives a displacement $D_1(\alpha)$ where $\alpha = \Omega_1 \tau /2$. Setting $\tau = \sqrt{2}\kappa/\omega \Omega_1$ so that $\alpha = \kappa/\sqrt{2}\omega$ displaces the mode from $Q_1=0$ to $Q_1=-\kappa/\omega$ because $D_1(\alpha) Q_1 D_1(\alpha)^\dagger = Q_1 - \sqrt{2} \Re\alpha = Q_1 - \kappa/\omega$. The qubit is first mapped into the SDF interaction basis ($\ket{+}_y$) with an $R_x(\pi/2)$ rotation, and is returned to $\ket{\downarrow}$ after the displacement with an $R_x(- \pi/2)$ rotation. The Rabi frequency of the SDF interaction was frequently recalibrated and on average we measured $\Omega_1= 2\pi\times \SI{2.23}{kHz}$.

\textit{Time evolution.} Two SDF interactions on B$_1$ and B$_2$ are applied during the time evolution. Their measured Rabi frequencies were, on average, $\sqrt{2}\kappa = 2\pi\times \SI{1.42}{kHz}$ and are calibrated within 2\% of each other. The duration $T$ of the geometric-phase dynamics is scaled according to the calibrated Rabi frequency.

\textit{Reconstruction measurement.} After the simulated time evolution, the system is in the entangled state $\ket{\Psi} = a_\downarrow \ket{\downarrow}\ket{\psi_\downarrow}_{B_1}\ket{\xi_\downarrow}_{B_2} + a_\uparrow \ket{\uparrow}\ket{\psi_\uparrow}_{B_1}\ket{\xi_\uparrow}_{B_2}$. In preparation of the reconstruction, a mid-circuit measurement projects the qubit state to either $\ket{\downarrow}$ or $\ket{\uparrow}$ through state-dependent fluorescence induced by a \SI{369.5}{nm} laser beam resonant with the \ce{^2S_{1/2}} $\ket{F=1}$ $\rightarrow$ \ce{^2P_{1/2}} $\ket{F=0}$ transition. The qubit states are inferred by thresholding the number of photons collected on an avalanche photodiode (measured state preparation and measurement fidelity of 99.5\%), and the outcomes of the measurement determine the probabilities $p_{\downarrow} = |a_\downarrow|^2$ and $p_{\uparrow} = |a_\uparrow|^2$. A measurement outcome of $\ket{\uparrow}$ induces significant decoherence of the bosonic modes due to photon recoils. Therefore, the reconstruction only proceeds if the measurement outcome is $\ket{\downarrow}$, for which no photon is emitted. Doing so projects the bosonic modes to $\ket{\psi_\downarrow}_{B_1}\ket{\xi_\downarrow}_{B_2}$. To retrieve $\ket{\psi_\uparrow}_{B_1}\ket{\xi_\uparrow}_{B_2}$ instead, we insert an $R_x(\pi)$ pulse that flips the qubit before the measurement.
After the mid-circuit measurement, the characteristic functions $\chi_{\downarrow}(i\beta_1, i\beta_2)$ and $\chi_{\uparrow}(i\beta_1, i\beta_2)$ corresponding to each qubit state are measured as described in the main text.
The full characteristic function is then the sum of both contributions, $\chi(i\beta_1,i\beta_2) = p_{\downarrow} \chi_{\downarrow}(i\beta_1, i\beta_2) + p_{\uparrow} \chi_{\uparrow}(i\beta_1, i\beta_2)$. The values $\beta_j= \Omega_2 \tau_j$ are scanned by varying the SDF-pulse duration $\tau_j$. The Rabi frequency was recalibrated between experiments and, on average, $\Omega_2 = 2\pi \times \SI{2.31}{kHz}$, resulting in combined pulse durations of up to $\SI{553}{\micro s}$.
We measured $\chi(i\beta_2)$ for $\beta_2\in [0, 5]$ and $\chi(i\beta_1, i\beta_2)$ for $\beta_1, \beta_2\in [0, 4]$ by varying the SDF pulse durations. Since the characteristic function is Hermitian, $\chi(i\beta_1, i\beta_2)^* = \chi(-i\beta_1, -i\beta_2)$, we used symmetry to find $\chi(i\beta_2)$ for $\beta_2<0$ and $\chi(i\beta_1, \beta_2)$ for $\beta_1<0$ or $\beta_2 < 0$. We did not measure the vanishing imaginary part of $\chi(i\beta_2)$ nor $\chi_\downarrow$ at $t=0$. The measured characteristic functions are shown in Extended Data Fig. 1.

\textit{Data acquisition.} The characteristic functions were measured in a way to average out the effects of drift. In each run of the experiment, we randomised the order of the displacement-pulse durations in which $\chi$ was reconstructed. For each run, the quantum circuit to obtain $\chi_{\downarrow}$ and $\chi_{\uparrow}$ was repeated until the mid-circuit measurement succeeded 500 times, resulting in 500 measurement repetitions of the reconstruction routine and 1000 measurements to obtain a value of the full $\chi$. Furthermore, the order of the displacements was randomised. Overall, each of the 1- and 2-dimensional experiments was repeated, respectively, four and two times and the results of the runs averaged for a total of 2000 and 1000 measurements for each duration. The bosonic mode frequencies were calibrated every \SI{6}{min}, while the full system parameters were recalibrated after the second experimental run. The 1-dimensional and the four 2-dimensional ($t=\{0, 0.9T, T, 2T\}$) experiments were done on five separate days with total durations of 15.6, 2.8, 8.8, 8.7 and 10.6 hours, respectively.

\textit{Noise sources} Decoherence of the bosonic modes, made up of motional heating and dephasing, was the dominant noise mechanism in our system. Motional heating was caused by electric field noise at the radial mode frequency, while motional dephasing arose from fluctuations in the harmonic trapping potential strength \cite{Wineland1998}. We measured the heating rate and the motional dephasing time of B$_1$ (representative for both modes) to be $\dot{\bar{n}} = \SI{0.2}{quanta~s^{-1}}$ and $T_2^* = \SI{35}{ms}$~\cite{MacDonell2022}. The motional coherence was limited by noise in the radio frequency (RF) trapping voltage, which was actively stabilised to mitigate fluctuations in the radial mode frequencies. Slow drifts of the trapping voltage were compensated for with frequent motional mode recalibrations.

\subsection*{Characteristic functions}

\begin{figure*}[t]
    \centering
    \includegraphics[width=\textwidth]{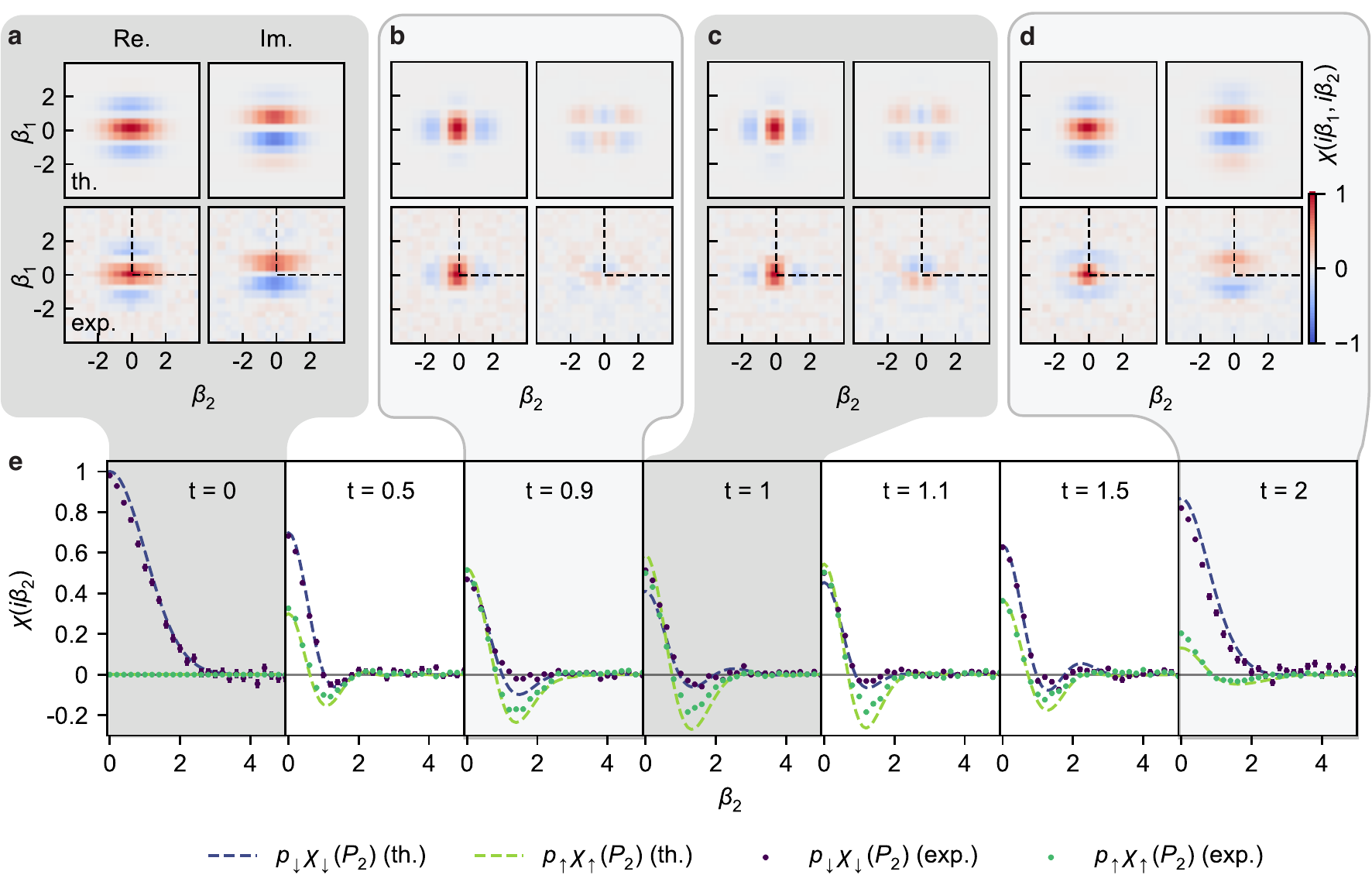}
    \caption{
    \textbf{Characteristic functions of the wavepacket measured for various evolution times.}
    \textbf{a-d}~Joint two-dimensional characteristic function $\chi(i\beta_1, i\beta_2) = p_\downarrow\chi_\downarrow(i\beta_1, i\beta_2) + p_\uparrow \chi_\uparrow(i\beta_1, i\beta_2)$ measured at times $t=\{0, 0.9T, T, 2T\}$ using the full pulse sequence of \cref{fig:experimental_pulse_sequence}. The real (left) and imaginary (right) parts are measured with and without an $R_x(\pi/2)$ pulse in the reconstruction. The top row shows theoretical predictions and the bottom experimental results. $\chi(i\beta_1, i\beta_2)$ were measured in the range $\beta_1, \beta_2 \in [0, 4]$ with $11\times 11$ equidistant samples (dashed quadrant). Values in the remaining three quadrants are obtained from the symmetry of $\chi(i\beta_1, i\beta_2)$.
    \textbf{e} One-dimensional characteristic functions $\chi_{\downarrow}(i\beta_2)$ and $\chi_{\uparrow}(i\beta_2)$ obtained by omitting displacements on B$_1$ in the reconstruction. $\beta_2$ was uniformly sampled in the range $[0, 5]$ with 26 points.
    Each two- and one-dimensional characteristic function was averaged over 1000 and 2000 measurements, respectively. Error bars in \textbf{e} represent one standard deviation based on quantum projection noise.}
    \label{fig:characteristic_func_probabilities}
\end{figure*}

\Cref{fig:characteristic_func_probabilities} shows the measured characteristic functions used to reconstruct the wavepacket probability densities. The two-dimensional characteristic functions (\cref{fig:characteristic_func_probabilities}a--d) require four measurements at each $t$ to obtain the real and imaginary parts of $p_\downarrow\chi_\downarrow(i\beta_1, i\beta_2)$ and $p_\uparrow\chi_\uparrow(i\beta_1, i\beta_2)$. The one-dimensional characteristic functions (\cref{fig:characteristic_func_probabilities}e) require two measurements to determine $p_\downarrow\chi_\downarrow(i\beta_2)$ and $p_\uparrow\chi_\uparrow(i\beta_2)$, as the vanishing imaginary part is not measured. In both one- and two-dimensional reconstructions, only positive values of $\beta_1$ and $\beta_2$ are sampled; the characteristic function in other ranges is obtained by symmetry.

We performed post-processing to remove artifacts associated with Fourier transformations between a characteristic function and its probability density. A non-zero DC offset appearing as background noise in the characteristic functions propagates into the probability densities at the origin \cite{Home2020}. Since background noise with a non-zero mean is a technical imperfection and is independent of the geometric-phase evolution, we correct for it in post-processing. 
We estimate the mean of the background noise by averaging $\chi(i\beta_1, i\beta_2)$ with $\sqrt{\beta_1^2+ \beta_1^2} \geq 3.6$ for the two-dimensional and $\chi(i\beta_2)$ with $\beta_2 \geq 3.6$ for the one-dimensional case, and offset the data by the negative of this average. This baseline correction was on average 0.02, with the largest correction of 0.03.

\subsection*{Phase coherence in the pulse sequence}

This appendix describes the experimental procedure to track the qubit and motional phases, ensuring phase coherence between sequential spin-motional interactions. 

The laser-induced excitations interacting with an ion with a qubit frequency $\omega_0$ and a motional mode with frequency $\omega_m$ are the carrier (c), red-sideband (rsb), and blue-sideband transitions (bsb). Their interaction Hamiltonians, after dropping high-frequency terms, are
\begin{align}
    H_\textrm{c} & = \frac{\Omega_c}{2}\sigma^+ e^{i \phi_c} e^{i(\omega_c - \omega_0)t} + \textrm{h.c.}, \\
    H_\textrm{rsb} & = \frac{\eta\Omega_r}{2}  \sigma^+ a e^{i \phi_r} e^{i (\omega_r - \omega_0 + \omega_m) t)} + \textrm{h.c.}, \\
    H_{\textrm{bsb}} & = \frac{\eta\Omega_b}{2}  \sigma^+ a^\dagger e^{i \phi_b} e^{i (\omega_b - \omega_0 - \omega_m) t)} + \textrm{h.c.},
\end{align}
where $\eta$ is the Lamb-Dicke parameter and $\Omega_{c,b,r}$ are the respective Rabi frequencies. $\omega_{c,b,r}$ and $\phi_{c,b,r}$ correspond, respectively, to the frequency differences and the phase differences of the two orthogonal Raman beams. Simultaneously driving the red- and blue-sidebands with $\Omega = \eta\Omega_r = \eta\Omega_b$ gives
\begin{alignat}{3}
    H_\textrm{SDF} & = H_\textrm{rsb} + && H_\textrm{bsb}   \nonumber \\
    & = \frac{  \Omega}{2} \sigma^+ [ && a e^{i \phi_r} e^{i (\omega_r - \omega_0 + \omega_m)t} + {} \nonumber \\
    & && a^\dagger e^{i \phi_b} e^{i (\omega_b - \omega_0 - \omega_m)t}  + \textrm{h.c.}].
    \label{eq:H_sdf_with_detunings}
\end{alignat}
We consider $\omega_r$ and $\omega_b$ to be set near-resonant with the red- and blue-sideband transitions,
\begin{align}
    \omega_r &= \omega_0 - \omega_m - \delta\omega_m + \delta\omega_0, \label{eq:rsb_frequency} \\ 
    \omega_b &= \omega_0 + \omega_m + \delta\omega_m + \delta\omega_0, \label{eq:bsb_frequency}
\end{align}
where $\delta\omega_0$ is an asymmetrical (center-line) detuning from the qubit frequency, and $\delta\omega_m$ is a symmetrical detuning from the motional mode frequency. With the spin phase $\phi_s = (\phi_r + \phi_b)/2$ and the motional phase $\phi_m = (\phi_b - \phi_r)/2$, \cref{eq:H_sdf_with_detunings} can be rewritten as 
\begin{align}
    H_\textrm{SDF} = \frac{\Omega}{2} \sigma^+ e^{ i(\delta \omega_0 t + \phi_s) } [ & a e^{-i( \delta \omega_m t + \phi_m)} + \nonumber\\ 
    & a^\dagger e^{i( \delta\omega_m t + \phi_m)} ] + \textrm{h.c.}
    \label{eq:SDF_Full}
\end{align}
This Hamiltonian corresponds to \cref{eq:H_sdf_main} in the main text by setting $\delta\omega_0 = 0$ and $\delta\omega_m = \delta$. The motional phase $\phi_m$ can be adjusted to selectively displace a mode along $Q$ or $P$. \Cref{eq:SDF_Full} shows that non-zero or miscalibrated $\delta\omega_0$ and $\delta\omega_m$ introduce a time-dependent phase offset to $\phi_s$ and $\phi_m$ which, if uncorrected, will lead to incorrect interactions. 

The qubit frequency detuning is, from \cref{eq:rsb_frequency} and \cref{eq:bsb_frequency}, $\delta\omega_0 = (\omega_b + \omega_r)/2 - \omega_0$. To avoid phase lags associated with $\phi_s$, we enforce $(\omega_b + \omega_r)/2 = \omega_c = \bar{\omega}_0$ for all pulses throughout the entire circuit, namely single-qubit rotations and SDF interactions on B$_1$ and B$_2$. Here, $\bar{\omega}_0$ indicates the qubit frequency measured via a Ramsey sequence in a separate calibration experiment. 

Likewise, \cref{eq:rsb_frequency} and \cref{eq:bsb_frequency} give the motional detuning as $\delta\omega_m = (\omega_b - \omega_r)/2 - \omega_m$. To avoid unwanted phase lags associated with $\phi_m$, we enforce $(\omega_b - \omega_r)/2 = \bar{\omega}_m$ for all SDF pulses throughout the circuit, where $\bar{\omega}_m$ is the experimentally measured motional frequency. There is an unavoidable phase lag due to the detuning $\delta$ required in the SDF interactions during the time evolution. To correct this, we add a motional phase offset of $\tau_1 \delta$ to the SDF interaction during the initial displacement, where $\tau_1$ is the duration of the initialisation. Furthermore, a motional phase offset of $(t + \tau_1)\delta$ is added to the reconstruction SDF pulses, where $t$ is the duration of the time evolution.

\subsection*{Calibration of motional frequencies}

\begin{figure}[t]
    \centering
    \includegraphics[]{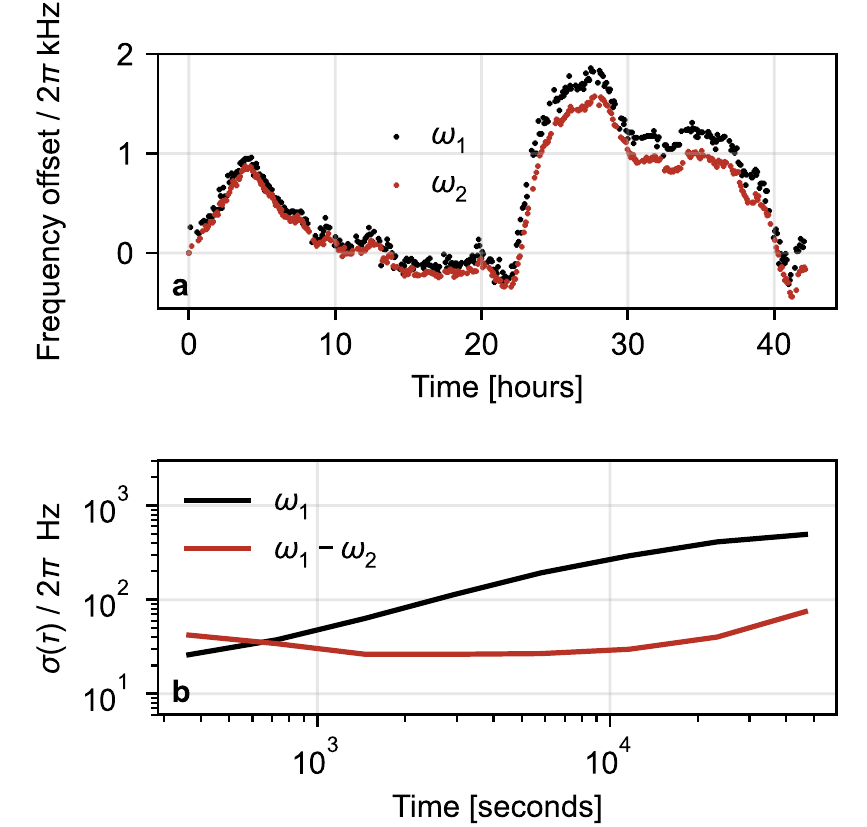}
    \caption{\textbf{Frequency drifts of radial motional modes.} 
    \textbf{a}~Time series of motional frequencies $\omega_{1,2}$ corresponding to B$_{1,2}$, measured using the calibration routine detailed in the text and plotted as the frequency offset from $\omega_1$ measured at $t = 0$. 
    \textbf{b}~Allan deviation of $\omega_1$, and of the difference between the two frequencies ($\omega_1 - \omega_2$).}
    \label{fig:mode_drift}
\end{figure}

We used a calibration scheduling routine to recalibrate parameters during each experiment and ensure high-fidelity implementations of the pulse sequence. Moreover, we optimised the scheduler to maximise the experimental duty cycle by analysing the temporal noise behaviours.

The data quality of the reconstructed densities depends on correctly setting the laser frequencies for the motional sideband interactions that enact SDF interactions. The motional frequencies $\omega_1$ and $\omega_2$ associated with B$_1$ and B$_2$ are calibrated as previously reported~\cite{MacDonell2022}. To do so, both SDF interactions are applied, but we set the fields associated with $\omega_2$ to be sufficiently off resonant while calibrating $\omega_1$. We prepare the state $\ket{\downarrow}\ket{0}_{\textrm{B}_1}$ and apply two sequential SDF pulses with a relative phase shift $\phi_m = \pi$. In the absence of frequency errors, the mode returns to its original state and a qubit measurement yields zero population in $\ket{\uparrow}$. However, in the presence of errors, the motion remains entangled with the qubit, giving a non-zero measured probability. The SDF fields' frequencies are then scanned, and a fit to the measurements yields the correct mode frequency. We repeat this procedure to calibrate $\omega_2$ by setting the SDF field associated with $\omega_1$ to be off resonant.

\Cref{fig:mode_drift} shows the drifts in the radial mode frequencies over time, which varied in a range of $2\pi \times \SI{2.2}{kHz}$ over 2 days. From numerical simulations, we determined that an error tolerance of about $10\%$ is required for the detuning ($\delta \simeq 2\pi\times \SI{667}{Hz}$) in the time evolution to obtain adequate results. Given that typical experiments lasted tens of hours, frequent recalibrations of the motional mode frequencies were necessary. To this end, we implemented a scheduling algorithm to interleave calibrations and experiments~\cite{Riesebos2021}. The scheduling rate was determined by choosing a time interval for which the Allan deviation was sufficiently small. From \cref{fig:mode_drift}b, we chose an interval of 6 minutes, corresponding to an Allan deviation of $2\pi \times \SI{26}{Hz}$ and satisfying the required tolerance. We also found highly correlated noise between the radial modes (see Extended Data Fig. 2), suggesting a common noise source (e.g., trap RF amplitude fluctuations). Therefore, to increase the experiment duty cycle, frequency offsets measured on B$_1$ were also used to correct for B$_2$.

\section*{Data Availability}

A repository containing data plotted in Fig. 3 and in Extended Data Fig. 1 is available at https://doi.org/10.5281/zenodo.7955887 (\cite{zenodo})


\begin{thebibliography}{55}%
\makeatletter
\providecommand \@ifxundefined [1]{%
 \@ifx{#1\undefined}
}%
\providecommand \@ifnum [1]{%
 \ifnum #1\expandafter \@firstoftwo
 \else \expandafter \@secondoftwo
 \fi
}%
\providecommand \@ifx [1]{%
 \ifx #1\expandafter \@firstoftwo
 \else \expandafter \@secondoftwo
 \fi
}%
\providecommand \natexlab [1]{#1}%
\providecommand \enquote  [1]{``#1''}%
\providecommand \bibnamefont  [1]{#1}%
\providecommand \bibfnamefont [1]{#1}%
\providecommand \citenamefont [1]{#1}%
\providecommand \href@noop [0]{\@secondoftwo}%
\providecommand \href [0]{\begingroup \@sanitize@url \@href}%
\providecommand \@href[1]{\@@startlink{#1}\@@href}%
\providecommand \@@href[1]{\endgroup#1\@@endlink}%
\providecommand \@sanitize@url [0]{\catcode `\\12\catcode `\$12\catcode
  `\&12\catcode `\#12\catcode `\^12\catcode `\_12\catcode `\%12\relax}%
\providecommand \@@startlink[1]{}%
\providecommand \@@endlink[0]{}%
\providecommand \url  [0]{\begingroup\@sanitize@url \@url }%
\providecommand \@url [1]{\endgroup\@href {#1}{\urlprefix }}%
\providecommand \urlprefix  [0]{URL }%
\providecommand \Eprint [0]{\href }%
\providecommand \doibase [0]{https://doi.org/}%
\providecommand \selectlanguage [0]{\@gobble}%
\providecommand \bibinfo  [0]{\@secondoftwo}%
\providecommand \bibfield  [0]{\@secondoftwo}%
\providecommand \translation [1]{[#1]}%
\providecommand \BibitemOpen [0]{}%
\providecommand \bibitemStop [0]{}%
\providecommand \bibitemNoStop [0]{.\EOS\space}%
\providecommand \EOS [0]{\spacefactor3000\relax}%
\providecommand \BibitemShut  [1]{\csname bibitem#1\endcsname}%
\let\auto@bib@innerbib\@empty
%</preamble>
\bibitem [{\citenamefont {Yarkony}(1996)}]{Yarkony}%
  \BibitemOpen
  \bibfield  {author} {\bibinfo {author} {\bibfnamefont {D.~R.}\ \bibnamefont
  {Yarkony}},\ }\bibfield  {title} {\bibinfo {title} {Diabolical conical
  intersections},\ }\href {https://doi.org/10.1103/RevModPhys.68.985}
  {\bibfield  {journal} {\bibinfo  {journal} {Rev. Mod. Phys.}\ }\textbf
  {\bibinfo {volume} {68}},\ \bibinfo {pages} {985} (\bibinfo {year}
  {1996})}\BibitemShut {NoStop}%
\bibitem [{\citenamefont {Domcke}\ \emph {et~al.}(2004)\citenamefont {Domcke},
  \citenamefont {Yarkony},\ and\ \citenamefont {Köppel}}]{Domcke}%
  \BibitemOpen
  \bibfield  {author} {\bibinfo {author} {\bibfnamefont {W.}~\bibnamefont
  {Domcke}}, \bibinfo {author} {\bibfnamefont {D.~R.}\ \bibnamefont
  {Yarkony}},\ and\ \bibinfo {author} {\bibfnamefont {H.}~\bibnamefont
  {Köppel}},\ }\href@noop {} {\emph {\bibinfo {title} {Conical Intersections:
  Electronic Structure, Dynamics and Spectroscopy}}}\ (\bibinfo  {publisher}
  {World Scientific Publishing},\ \bibinfo {year} {2004})\BibitemShut {NoStop}%
\bibitem [{\citenamefont {Larson}\ \emph {et~al.}(2020)\citenamefont {Larson},
  \citenamefont {Sjöqvist},\ and\ \citenamefont {Öhberg}}]{LarsonBook}%
  \BibitemOpen
  \bibfield  {author} {\bibinfo {author} {\bibfnamefont {J.}~\bibnamefont
  {Larson}}, \bibinfo {author} {\bibfnamefont {E.}~\bibnamefont {Sjöqvist}},\
  and\ \bibinfo {author} {\bibfnamefont {P.}~\bibnamefont {Öhberg}},\
  }\href@noop {} {\emph {\bibinfo {title} {Conical Intersections in Physics}}}\
  (\bibinfo  {publisher} {Springer Cham},\ \bibinfo {year} {2020})\BibitemShut
  {NoStop}%
\bibitem [{\citenamefont {Domcke}\ and\ \citenamefont
  {Yarkony}(2012)}]{Domcke2}%
  \BibitemOpen
  \bibfield  {author} {\bibinfo {author} {\bibfnamefont {W.}~\bibnamefont
  {Domcke}}\ and\ \bibinfo {author} {\bibfnamefont {D.~R.}\ \bibnamefont
  {Yarkony}},\ }\bibfield  {title} {\bibinfo {title} {Role of conical
  intersections in molecular spectroscopy and photoinduced chemical dynamics},\
  }\href {https://doi.org/10.1146/annurev-physchem-032210-103522} {\bibfield
  {journal} {\bibinfo  {journal} {Annu. Rev. Phys. Chem.}\ }\textbf {\bibinfo
  {volume} {63}},\ \bibinfo {pages} {325} (\bibinfo {year} {2012})}\BibitemShut
  {NoStop}%
\bibitem [{\citenamefont {Berry}(1984)}]{Berry}%
  \BibitemOpen
  \bibfield  {author} {\bibinfo {author} {\bibfnamefont {M.~V.}\ \bibnamefont
  {Berry}},\ }\bibfield  {title} {\bibinfo {title} {Quantal phase factors
  accompanying adiabatic changes},\ }\href
  {https://doi.org/10.1098/rspa.1984.0023} {\bibfield  {journal} {\bibinfo
  {journal} {Proc. R. Soc. Lond. A}\ }\textbf {\bibinfo {volume} {392}},\
  \bibinfo {pages} {47} (\bibinfo {year} {1984})}\BibitemShut {NoStop}%
\bibitem [{\citenamefont {Longuet-Higgins}\ \emph {et~al.}(1958)\citenamefont
  {Longuet-Higgins}, \citenamefont {Öpik}, \citenamefont {Pryce},\ and\
  \citenamefont {Sack}}]{Longuet}%
  \BibitemOpen
  \bibfield  {author} {\bibinfo {author} {\bibfnamefont {H.~C.}\ \bibnamefont
  {Longuet-Higgins}}, \bibinfo {author} {\bibfnamefont {U.}~\bibnamefont
  {Öpik}}, \bibinfo {author} {\bibfnamefont {M.~H.~L.}\ \bibnamefont
  {Pryce}},\ and\ \bibinfo {author} {\bibfnamefont {R.~A.}\ \bibnamefont
  {Sack}},\ }\bibfield  {title} {\bibinfo {title} {Studies of the
  {J}ahn-{T}eller effect {II.} {T}he dynamical problem},\ }\href
  {https://doi.org/10.1098/rspa.1958.0022} {\bibfield  {journal} {\bibinfo
  {journal} {Proc. R. Soc. Lond. A}\ }\textbf {\bibinfo {volume} {244}},\
  \bibinfo {pages} {1} (\bibinfo {year} {1958})}\BibitemShut {NoStop}%
\bibitem [{\citenamefont {Mead}\ and\ \citenamefont
  {Truhlar}(1979)}]{Mead-Truhlar}%
  \BibitemOpen
  \bibfield  {author} {\bibinfo {author} {\bibfnamefont {C.~A.}\ \bibnamefont
  {Mead}}\ and\ \bibinfo {author} {\bibfnamefont {D.~G.}\ \bibnamefont
  {Truhlar}},\ }\bibfield  {title} {\bibinfo {title} {On the determination of
  {B}orn–{O}ppenheimer nuclear motion wave functions including complications
  due to conical intersections and identical nuclei},\ }\href
  {https://doi.org/10.1063/1.437734} {\bibfield  {journal} {\bibinfo  {journal}
  {J. Chem. Phys.}\ }\textbf {\bibinfo {volume} {70}},\ \bibinfo {pages} {2284}
  (\bibinfo {year} {1979})}\BibitemShut {NoStop}%
\bibitem [{\citenamefont {Schön}\ and\ \citenamefont {Köppel}(1995)}]{Schon}%
  \BibitemOpen
  \bibfield  {author} {\bibinfo {author} {\bibfnamefont {J.}~\bibnamefont
  {Schön}}\ and\ \bibinfo {author} {\bibfnamefont {H.}~\bibnamefont
  {Köppel}},\ }\bibfield  {title} {\bibinfo {title} {{Geometric phase effects
  and wave packet dynamics on intersecting potential energy surfaces}},\ }\href
  {https://doi.org/10.1063/1.469988} {\bibfield  {journal} {\bibinfo  {journal}
  {J. Chem. Phys.}\ }\textbf {\bibinfo {volume} {103}},\ \bibinfo {pages}
  {9292} (\bibinfo {year} {1995})}\BibitemShut {NoStop}%
\bibitem [{\citenamefont {Ryabinkin}\ \emph {et~al.}(2017)\citenamefont
  {Ryabinkin}, \citenamefont {Joubert-Doriol},\ and\ \citenamefont
  {Izmaylov}}]{Ryabinkin}%
  \BibitemOpen
  \bibfield  {author} {\bibinfo {author} {\bibfnamefont {I.~G.}\ \bibnamefont
  {Ryabinkin}}, \bibinfo {author} {\bibfnamefont {L.}~\bibnamefont
  {Joubert-Doriol}},\ and\ \bibinfo {author} {\bibfnamefont {A.~F.}\
  \bibnamefont {Izmaylov}},\ }\bibfield  {title} {\bibinfo {title} {Geometric
  phase effects in nonadiabatic dynamics near conical intersections},\ }\href
  {https://doi.org/10.1021/acs.accounts.7b00220} {\bibfield  {journal}
  {\bibinfo  {journal} {Acc. Chem. Res.}\ }\textbf {\bibinfo {volume} {50}},\
  \bibinfo {pages} {1785} (\bibinfo {year} {2017})}\BibitemShut {NoStop}%
\bibitem [{\citenamefont {Mead}(1980)}]{Mead1980}%
  \BibitemOpen
  \bibfield  {author} {\bibinfo {author} {\bibfnamefont {C.~A.}\ \bibnamefont
  {Mead}},\ }\bibfield  {title} {\bibinfo {title} {Superposition of reactive
  and nonreactive scattering amplitudes in the presence of a conical
  intersection},\ }\href {https://doi.org/10.1063/1.439600} {\bibfield
  {journal} {\bibinfo  {journal} {J. Chem. Phys.}\ }\textbf {\bibinfo {volume}
  {72}},\ \bibinfo {pages} {3839} (\bibinfo {year} {1980})}\BibitemShut
  {NoStop}%
\bibitem [{\citenamefont {Lepetit}\ and\ \citenamefont
  {Kuppermann}(1990)}]{Lepetit1990}%
  \BibitemOpen
  \bibfield  {author} {\bibinfo {author} {\bibfnamefont {B.}~\bibnamefont
  {Lepetit}}\ and\ \bibinfo {author} {\bibfnamefont {A.}~\bibnamefont
  {Kuppermann}},\ }\bibfield  {title} {\bibinfo {title} {Numerical study of the
  geometric phase in the \ce{H + H2} reaction},\ }\href
  {https://doi.org/https://doi.org/10.1016/0009-2614(90)87154-J} {\bibfield
  {journal} {\bibinfo  {journal} {Chem. Phys. Lett.}\ }\textbf {\bibinfo
  {volume} {166}},\ \bibinfo {pages} {581} (\bibinfo {year}
  {1990})}\BibitemShut {NoStop}%
\bibitem [{\citenamefont {Althorpe}(2006)}]{Althorpe2006}%
  \BibitemOpen
  \bibfield  {author} {\bibinfo {author} {\bibfnamefont {S.~C.}\ \bibnamefont
  {Althorpe}},\ }\bibfield  {title} {\bibinfo {title} {General explanation of
  geometric phase effects in reactive systems: Unwinding the nuclear wave
  function using simple topology},\ }\href {https://doi.org/10.1063/1.2161220}
  {\bibfield  {journal} {\bibinfo  {journal} {J. Chem. Phys.}\ }\textbf
  {\bibinfo {volume} {124}},\ \bibinfo {pages} {084105} (\bibinfo {year}
  {2006})}\BibitemShut {NoStop}%
\bibitem [{\citenamefont {Althorpe}\ \emph {et~al.}(2008)\citenamefont
  {Althorpe}, \citenamefont {Stecher},\ and\ \citenamefont
  {Bouakline}}]{Althorpe2008}%
  \BibitemOpen
  \bibfield  {author} {\bibinfo {author} {\bibfnamefont {S.~C.}\ \bibnamefont
  {Althorpe}}, \bibinfo {author} {\bibfnamefont {T.}~\bibnamefont {Stecher}},\
  and\ \bibinfo {author} {\bibfnamefont {F.}~\bibnamefont {Bouakline}},\
  }\bibfield  {title} {\bibinfo {title} {Effect of the geometric phase on
  nuclear dynamics at a conical intersection: Extension of a recent topological
  approach from one to two coupled surfaces},\ }\href
  {https://doi.org/10.1063/1.3031215} {\bibfield  {journal} {\bibinfo
  {journal} {J. Chem. Phys.}\ }\textbf {\bibinfo {volume} {129}},\ \bibinfo
  {pages} {214117} (\bibinfo {year} {2008})}\BibitemShut {NoStop}%
\bibitem [{\citenamefont {Kendrick}(1997)}]{Kendrick1997}%
  \BibitemOpen
  \bibfield  {author} {\bibinfo {author} {\bibfnamefont {B.}~\bibnamefont
  {Kendrick}},\ }\bibfield  {title} {\bibinfo {title} {Geometric phase effects
  in the vibrational spectrum of \ce{Na3(X)}},\ }\href
  {https://doi.org/10.1103/PhysRevLett.79.2431} {\bibfield  {journal} {\bibinfo
   {journal} {Phys. Rev. Lett.}\ }\textbf {\bibinfo {volume} {79}},\ \bibinfo
  {pages} {2431} (\bibinfo {year} {1997})}\BibitemShut {NoStop}%
\bibitem [{\citenamefont {Applegate}\ \emph {et~al.}(2003)\citenamefont
  {Applegate}, \citenamefont {Barckholtz},\ and\ \citenamefont
  {Miller}}]{Applegate}%
  \BibitemOpen
  \bibfield  {author} {\bibinfo {author} {\bibfnamefont {B.~E.}\ \bibnamefont
  {Applegate}}, \bibinfo {author} {\bibfnamefont {T.~A.}\ \bibnamefont
  {Barckholtz}},\ and\ \bibinfo {author} {\bibfnamefont {T.~A.}\ \bibnamefont
  {Miller}},\ }\bibfield  {title} {\bibinfo {title} {Explorations of conical
  intersections and their ramifications for chemistry through the
  {J}ahn–{T}eller effect},\ }\href {https://doi.org/10.1039/A910269H}
  {\bibfield  {journal} {\bibinfo  {journal} {Chem. Soc. Rev.}\ }\textbf
  {\bibinfo {volume} {32}},\ \bibinfo {pages} {38} (\bibinfo {year}
  {2003})}\BibitemShut {NoStop}%
\bibitem [{\citenamefont {Englman}(2016)}]{Englman}%
  \BibitemOpen
  \bibfield  {author} {\bibinfo {author} {\bibfnamefont {R.}~\bibnamefont
  {Englman}},\ }\bibfield  {title} {\bibinfo {title} {Spectroscopic
  detectability of the molecular {A}haronov-{B}ohm effect},\ }\href
  {https://doi.org/10.1063/1.4939243} {\bibfield  {journal} {\bibinfo
  {journal} {J. Chem. Phys.}\ }\textbf {\bibinfo {volume} {144}},\ \bibinfo
  {pages} {024103} (\bibinfo {year} {2016})}\BibitemShut {NoStop}%
\bibitem [{\citenamefont {Yuan}\ \emph {et~al.}(2018)\citenamefont {Yuan},
  \citenamefont {Guan}, \citenamefont {Chen}, \citenamefont {Zhao},
  \citenamefont {Yu}, \citenamefont {Luo}, \citenamefont {Tan}, \citenamefont
  {Xie}, \citenamefont {Wang}, \citenamefont {Sun}, \citenamefont {Zhang},\
  and\ \citenamefont {Yang}}]{Daofu}%
  \BibitemOpen
  \bibfield  {author} {\bibinfo {author} {\bibfnamefont {D.}~\bibnamefont
  {Yuan}}, \bibinfo {author} {\bibfnamefont {Y.}~\bibnamefont {Guan}}, \bibinfo
  {author} {\bibfnamefont {W.}~\bibnamefont {Chen}}, \bibinfo {author}
  {\bibfnamefont {H.}~\bibnamefont {Zhao}}, \bibinfo {author} {\bibfnamefont
  {S.}~\bibnamefont {Yu}}, \bibinfo {author} {\bibfnamefont {C.}~\bibnamefont
  {Luo}}, \bibinfo {author} {\bibfnamefont {Y.}~\bibnamefont {Tan}}, \bibinfo
  {author} {\bibfnamefont {T.}~\bibnamefont {Xie}}, \bibinfo {author}
  {\bibfnamefont {X.}~\bibnamefont {Wang}}, \bibinfo {author} {\bibfnamefont
  {Z.}~\bibnamefont {Sun}}, \bibinfo {author} {\bibfnamefont {D.~H.}\
  \bibnamefont {Zhang}},\ and\ \bibinfo {author} {\bibfnamefont
  {X.}~\bibnamefont {Yang}},\ }\bibfield  {title} {\bibinfo {title}
  {Observation of the geometric phase effect in the \ce{H + HD \to H2 + D}
  reaction},\ }\href {https://doi.org/10.1126/science.aav1356} {\bibfield
  {journal} {\bibinfo  {journal} {Science}\ }\textbf {\bibinfo {volume}
  {362}},\ \bibinfo {pages} {1289} (\bibinfo {year} {2018})}\BibitemShut
  {NoStop}%
\bibitem [{\citenamefont {Yuan}\ \emph {et~al.}(2020)\citenamefont {Yuan},
  \citenamefont {Huang}, \citenamefont {Chen}, \citenamefont {Zhao},
  \citenamefont {Yu}, \citenamefont {Luo}, \citenamefont {Tan}, \citenamefont
  {Wang}, \citenamefont {Wang}, \citenamefont {Sun},\ and\ \citenamefont
  {Yang}}]{DaofuII}%
  \BibitemOpen
  \bibfield  {author} {\bibinfo {author} {\bibfnamefont {D.}~\bibnamefont
  {Yuan}}, \bibinfo {author} {\bibfnamefont {Y.}~\bibnamefont {Huang}},
  \bibinfo {author} {\bibfnamefont {W.}~\bibnamefont {Chen}}, \bibinfo {author}
  {\bibfnamefont {H.}~\bibnamefont {Zhao}}, \bibinfo {author} {\bibfnamefont
  {S.}~\bibnamefont {Yu}}, \bibinfo {author} {\bibfnamefont {C.}~\bibnamefont
  {Luo}}, \bibinfo {author} {\bibfnamefont {Y.}~\bibnamefont {Tan}}, \bibinfo
  {author} {\bibfnamefont {S.}~\bibnamefont {Wang}}, \bibinfo {author}
  {\bibfnamefont {X.}~\bibnamefont {Wang}}, \bibinfo {author} {\bibfnamefont
  {Z.}~\bibnamefont {Sun}},\ and\ \bibinfo {author} {\bibfnamefont
  {X.}~\bibnamefont {Yang}},\ }\bibfield  {title} {\bibinfo {title}
  {Observation of the geometric phase effect in the \ce{H + HD \to H2 + D}
  reaction below the conical intersection},\ }\href
  {https://doi.org/10.1038/s41467-020-17381-4} {\bibfield  {journal} {\bibinfo
  {journal} {Nat. Commun.}\ }\textbf {\bibinfo {volume} {11}},\ \bibinfo
  {pages} {3640} (\bibinfo {year} {2020})}\BibitemShut {NoStop}%
\bibitem [{\citenamefont {Cina}\ and\ \citenamefont
  {Romero‐Rochin}(1990)}]{Cina1990}%
  \BibitemOpen
  \bibfield  {author} {\bibinfo {author} {\bibfnamefont {J.~A.}\ \bibnamefont
  {Cina}}\ and\ \bibinfo {author} {\bibfnamefont {V.}~\bibnamefont
  {Romero‐Rochin}},\ }\bibfield  {title} {\bibinfo {title} {Optical impulsive
  excitation of molecular pseudorotation in {J}ahn–{T}eller systems},\ }\href
  {https://doi.org/10.1063/1.458769} {\bibfield  {journal} {\bibinfo  {journal}
  {J. Chem. Phys.}\ }\textbf {\bibinfo {volume} {93}},\ \bibinfo {pages} {3844}
  (\bibinfo {year} {1990})}\BibitemShut {NoStop}%
\bibitem [{\citenamefont {Cina}(1991)}]{Cina1991}%
  \BibitemOpen
  \bibfield  {author} {\bibinfo {author} {\bibfnamefont {J.~A.}\ \bibnamefont
  {Cina}},\ }\bibfield  {title} {\bibinfo {title} {Phase-controlled optical
  pulses and the adiabatic electronic sign change},\ }\href
  {https://doi.org/10.1103/PhysRevLett.66.1146} {\bibfield  {journal} {\bibinfo
   {journal} {Phys. Rev. Lett.}\ }\textbf {\bibinfo {volume} {66}},\ \bibinfo
  {pages} {1146} (\bibinfo {year} {1991})}\BibitemShut {NoStop}%
\bibitem [{\citenamefont {Cina}\ \emph {et~al.}(1993)\citenamefont {Cina},
  \citenamefont {T.~J.~Smith},\ and\ \citenamefont {Romero-Rochin}}]{Cina1993}%
  \BibitemOpen
  \bibfield  {author} {\bibinfo {author} {\bibfnamefont {J.~A.}\ \bibnamefont
  {Cina}}, \bibinfo {author} {\bibfnamefont {J.}~\bibnamefont {T.~J.~Smith}},\
  and\ \bibinfo {author} {\bibfnamefont {V.}~\bibnamefont {Romero-Rochin}},\
  }\bibfield  {title} {\bibinfo {title} {Time-resolved optical tests for
  electronic geometric phase development},\ }\href@noop {} {\bibfield
  {journal} {\bibinfo  {journal} {Adv. in Chem. Phys.}\ }\textbf {\bibinfo
  {volume} {83}},\ \bibinfo {pages} {1} (\bibinfo {year} {1993})}\BibitemShut
  {NoStop}%
\bibitem [{\citenamefont {Castro~Neto}\ \emph {et~al.}(2009)\citenamefont
  {Castro~Neto}, \citenamefont {Guinea}, \citenamefont {Peres}, \citenamefont
  {Novoselov},\ and\ \citenamefont {Geim}}]{Castro}%
  \BibitemOpen
  \bibfield  {author} {\bibinfo {author} {\bibfnamefont {A.~H.}\ \bibnamefont
  {Castro~Neto}}, \bibinfo {author} {\bibfnamefont {F.}~\bibnamefont {Guinea}},
  \bibinfo {author} {\bibfnamefont {N.~M.~R.}\ \bibnamefont {Peres}}, \bibinfo
  {author} {\bibfnamefont {K.~S.}\ \bibnamefont {Novoselov}},\ and\ \bibinfo
  {author} {\bibfnamefont {A.~K.}\ \bibnamefont {Geim}},\ }\bibfield  {title}
  {\bibinfo {title} {The electronic properties of graphene},\ }\href
  {https://doi.org/10.1103/RevModPhys.81.109} {\bibfield  {journal} {\bibinfo
  {journal} {Rev. Mod. Phys.}\ }\textbf {\bibinfo {volume} {81}},\ \bibinfo
  {pages} {109} (\bibinfo {year} {2009})}\BibitemShut {NoStop}%
\bibitem [{\citenamefont {Ran}\ \emph {et~al.}(2009)\citenamefont {Ran},
  \citenamefont {Wang}, \citenamefont {Zhai}, \citenamefont {Vishwanath},\ and\
  \citenamefont {Lee}}]{Ran_Superconductor}%
  \BibitemOpen
  \bibfield  {author} {\bibinfo {author} {\bibfnamefont {Y.}~\bibnamefont
  {Ran}}, \bibinfo {author} {\bibfnamefont {F.}~\bibnamefont {Wang}}, \bibinfo
  {author} {\bibfnamefont {H.}~\bibnamefont {Zhai}}, \bibinfo {author}
  {\bibfnamefont {A.}~\bibnamefont {Vishwanath}},\ and\ \bibinfo {author}
  {\bibfnamefont {D.-H.}\ \bibnamefont {Lee}},\ }\bibfield  {title} {\bibinfo
  {title} {Nodal spin density wave and band topology of the {FeAs}-based
  materials},\ }\href {https://doi.org/10.1103/PhysRevB.79.014505} {\bibfield
  {journal} {\bibinfo  {journal} {Phys. Rev. B}\ }\textbf {\bibinfo {volume}
  {79}},\ \bibinfo {pages} {014505} (\bibinfo {year} {2009})}\BibitemShut
  {NoStop}%
\bibitem [{\citenamefont {Rashba}(1959)}]{Rashba}%
  \BibitemOpen
  \bibfield  {author} {\bibinfo {author} {\bibfnamefont {E.}~\bibnamefont
  {Rashba}},\ }\bibfield  {title} {\bibinfo {title} {Symmetry of energy bands
  in crystals of wurtzite type: {I.} symmetry of bands disregarding spin-orbit
  interaction},\ }\href@noop {} {\bibfield  {journal} {\bibinfo  {journal}
  {Sov. Phys.-Solid State}\ }\textbf {\bibinfo {volume} {1}},\ \bibinfo {pages}
  {368} (\bibinfo {year} {1959})}\BibitemShut {NoStop}%
\bibitem [{\citenamefont {Dresselhaus}(1955)}]{Dresselhaus}%
  \BibitemOpen
  \bibfield  {author} {\bibinfo {author} {\bibfnamefont {G.}~\bibnamefont
  {Dresselhaus}},\ }\bibfield  {title} {\bibinfo {title} {Spin-orbit coupling
  effects in zinc blende structures},\ }\href
  {https://doi.org/10.1103/PhysRev.100.580} {\bibfield  {journal} {\bibinfo
  {journal} {Phys. Rev.}\ }\textbf {\bibinfo {volume} {100}},\ \bibinfo {pages}
  {580} (\bibinfo {year} {1955})}\BibitemShut {NoStop}%
\bibitem [{\citenamefont {Cina}(2008)}]{Cina2008}%
  \BibitemOpen
  \bibfield  {author} {\bibinfo {author} {\bibfnamefont {J.~A.}\ \bibnamefont
  {Cina}},\ }\bibfield  {title} {\bibinfo {title} {Wave-packet interferometry
  and molecular state reconstruction: Spectroscopic adventures on the left-hand
  side of the {S}chrödinger equation},\ }\href
  {https://doi.org/10.1146/annurev.physchem.59.032607.093753} {\bibfield
  {journal} {\bibinfo  {journal} {Annu. Rev. Phys. Chem.}\ }\textbf {\bibinfo
  {volume} {59}},\ \bibinfo {pages} {319} (\bibinfo {year} {2008})}\BibitemShut
  {NoStop}%
\bibitem [{\citenamefont {Buluta}\ and\ \citenamefont {Nori}(2009)}]{Buluta}%
  \BibitemOpen
  \bibfield  {author} {\bibinfo {author} {\bibfnamefont {I.}~\bibnamefont
  {Buluta}}\ and\ \bibinfo {author} {\bibfnamefont {F.}~\bibnamefont {Nori}},\
  }\bibfield  {title} {\bibinfo {title} {Quantum simulators},\ }\href
  {https://doi.org/10.1126/science.1177838} {\bibfield  {journal} {\bibinfo
  {journal} {Science}\ }\textbf {\bibinfo {volume} {326}},\ \bibinfo {pages}
  {108} (\bibinfo {year} {2009})}\BibitemShut {NoStop}%
\bibitem [{\citenamefont {Blatt}\ and\ \citenamefont {Roos}(2012)}]{Roos}%
  \BibitemOpen
  \bibfield  {author} {\bibinfo {author} {\bibfnamefont {R.}~\bibnamefont
  {Blatt}}\ and\ \bibinfo {author} {\bibfnamefont {C.~F.}\ \bibnamefont
  {Roos}},\ }\bibfield  {title} {\bibinfo {title} {Quantum simulations with
  trapped ions},\ }\href {https://doi.org/10.1038/nphys2252} {\bibfield
  {journal} {\bibinfo  {journal} {Nat. Phys.}\ }\textbf {\bibinfo {volume}
  {8}},\ \bibinfo {pages} {277} (\bibinfo {year} {2012})}\BibitemShut {NoStop}%
\bibitem [{\citenamefont {Aspuru-Guzik}\ and\ \citenamefont
  {Walther}(2012)}]{Aspuru}%
  \BibitemOpen
  \bibfield  {author} {\bibinfo {author} {\bibfnamefont {A.}~\bibnamefont
  {Aspuru-Guzik}}\ and\ \bibinfo {author} {\bibfnamefont {P.}~\bibnamefont
  {Walther}},\ }\bibfield  {title} {\bibinfo {title} {Photonic quantum
  simulators},\ }\href {https://doi.org/10.1038/nphys2253} {\bibfield
  {journal} {\bibinfo  {journal} {Nat. Phys.}\ }\textbf {\bibinfo {volume}
  {8}},\ \bibinfo {pages} {285} (\bibinfo {year} {2012})}\BibitemShut {NoStop}%
\bibitem [{\citenamefont {McArdle}\ \emph {et~al.}(2020)\citenamefont
  {McArdle}, \citenamefont {Endo}, \citenamefont {Aspuru-Guzik}, \citenamefont
  {Benjamin},\ and\ \citenamefont {Yuan}}]{McArdle}%
  \BibitemOpen
  \bibfield  {author} {\bibinfo {author} {\bibfnamefont {S.}~\bibnamefont
  {McArdle}}, \bibinfo {author} {\bibfnamefont {S.}~\bibnamefont {Endo}},
  \bibinfo {author} {\bibfnamefont {A.}~\bibnamefont {Aspuru-Guzik}}, \bibinfo
  {author} {\bibfnamefont {S.~C.}\ \bibnamefont {Benjamin}},\ and\ \bibinfo
  {author} {\bibfnamefont {X.}~\bibnamefont {Yuan}},\ }\bibfield  {title}
  {\bibinfo {title} {Quantum computational chemistry},\ }\href
  {https://doi.org/10.1103/RevModPhys.92.015003} {\bibfield  {journal}
  {\bibinfo  {journal} {Rev. Mod. Phys.}\ }\textbf {\bibinfo {volume} {92}},\
  \bibinfo {pages} {015003} (\bibinfo {year} {2020})}\BibitemShut {NoStop}%
\bibitem [{\citenamefont {Gorman}\ \emph {et~al.}(2018)\citenamefont {Gorman},
  \citenamefont {Hemmerling}, \citenamefont {Megidish}, \citenamefont
  {Moeller}, \citenamefont {Schindler}, \citenamefont {Sarovar},\ and\
  \citenamefont {Haeffner}}]{Haeffner2018}%
  \BibitemOpen
  \bibfield  {author} {\bibinfo {author} {\bibfnamefont {J.~D.}\ \bibnamefont
  {Gorman}}, \bibinfo {author} {\bibfnamefont {B.}~\bibnamefont {Hemmerling}},
  \bibinfo {author} {\bibfnamefont {E.}~\bibnamefont {Megidish}}, \bibinfo
  {author} {\bibfnamefont {S.~A.}\ \bibnamefont {Moeller}}, \bibinfo {author}
  {\bibfnamefont {P.}~\bibnamefont {Schindler}}, \bibinfo {author}
  {\bibfnamefont {M.}~\bibnamefont {Sarovar}},\ and\ \bibinfo {author}
  {\bibfnamefont {H.}~\bibnamefont {Haeffner}},\ }\bibfield  {title} {\bibinfo
  {title} {Engineering vibrationally assisted energy transfer in a trapped-ion
  quantum simulator},\ }\href {https://doi.org/10.1103/PhysRevX.8.011038}
  {\bibfield  {journal} {\bibinfo  {journal} {Phys. Rev. X}\ }\textbf {\bibinfo
  {volume} {8}},\ \bibinfo {pages} {011038} (\bibinfo {year}
  {2018})}\BibitemShut {NoStop}%
\bibitem [{\citenamefont {Duca}\ \emph {et~al.}(2015)\citenamefont {Duca},
  \citenamefont {Li}, \citenamefont {Reitter}, \citenamefont {Bloch},
  \citenamefont {Schleier-Smith},\ and\ \citenamefont {Schneider}}]{Bloch2015}%
  \BibitemOpen
  \bibfield  {author} {\bibinfo {author} {\bibfnamefont {L.}~\bibnamefont
  {Duca}}, \bibinfo {author} {\bibfnamefont {T.}~\bibnamefont {Li}}, \bibinfo
  {author} {\bibfnamefont {M.}~\bibnamefont {Reitter}}, \bibinfo {author}
  {\bibfnamefont {I.}~\bibnamefont {Bloch}}, \bibinfo {author} {\bibfnamefont
  {M.}~\bibnamefont {Schleier-Smith}},\ and\ \bibinfo {author} {\bibfnamefont
  {U.}~\bibnamefont {Schneider}},\ }\bibfield  {title} {\bibinfo {title} {An
  {A}haronov-{B}ohm interferometer for determining {B}loch band topology},\
  }\href {https://doi.org/10.1126/science.1259052} {\bibfield  {journal}
  {\bibinfo  {journal} {Science}\ }\textbf {\bibinfo {volume} {347}},\ \bibinfo
  {pages} {288} (\bibinfo {year} {2015})}\BibitemShut {NoStop}%
\bibitem [{\citenamefont {Brown}\ \emph {et~al.}(2022)\citenamefont {Brown},
  \citenamefont {Chang}, \citenamefont {Schwarz}, \citenamefont {Leung},
  \citenamefont {Kozii}, \citenamefont {Avdoshkin}, \citenamefont {Moore},\
  and\ \citenamefont {Stamper-Kurn}}]{Brown2022}%
  \BibitemOpen
  \bibfield  {author} {\bibinfo {author} {\bibfnamefont {C.~D.}\ \bibnamefont
  {Brown}}, \bibinfo {author} {\bibfnamefont {S.-W.}\ \bibnamefont {Chang}},
  \bibinfo {author} {\bibfnamefont {M.~N.}\ \bibnamefont {Schwarz}}, \bibinfo
  {author} {\bibfnamefont {T.-H.}\ \bibnamefont {Leung}}, \bibinfo {author}
  {\bibfnamefont {V.}~\bibnamefont {Kozii}}, \bibinfo {author} {\bibfnamefont
  {A.}~\bibnamefont {Avdoshkin}}, \bibinfo {author} {\bibfnamefont {J.~E.}\
  \bibnamefont {Moore}},\ and\ \bibinfo {author} {\bibfnamefont
  {D.}~\bibnamefont {Stamper-Kurn}},\ }\bibfield  {title} {\bibinfo {title}
  {Direct geometric probe of singularities in band structure},\ }\href
  {https://doi.org/10.1126/science.abm6442} {\bibfield  {journal} {\bibinfo
  {journal} {Science}\ }\textbf {\bibinfo {volume} {377}},\ \bibinfo {pages}
  {1319} (\bibinfo {year} {2022})}\BibitemShut {NoStop}%
\bibitem [{\citenamefont {Gambetta}\ \emph {et~al.}(2021)\citenamefont
  {Gambetta}, \citenamefont {Zhang}, \citenamefont {Hennrich}, \citenamefont
  {Lesanovsky},\ and\ \citenamefont {Li}}]{Gambetta}%
  \BibitemOpen
  \bibfield  {author} {\bibinfo {author} {\bibfnamefont {F.~M.}\ \bibnamefont
  {Gambetta}}, \bibinfo {author} {\bibfnamefont {C.}~\bibnamefont {Zhang}},
  \bibinfo {author} {\bibfnamefont {M.}~\bibnamefont {Hennrich}}, \bibinfo
  {author} {\bibfnamefont {I.}~\bibnamefont {Lesanovsky}},\ and\ \bibinfo
  {author} {\bibfnamefont {W.}~\bibnamefont {Li}},\ }\bibfield  {title}
  {\bibinfo {title} {Exploring the many-body dynamics near a conical
  intersection with trapped {R}ydberg ions},\ }\href
  {https://doi.org/10.1103/PhysRevLett.126.233404} {\bibfield  {journal}
  {\bibinfo  {journal} {Phys. Rev. Lett.}\ }\textbf {\bibinfo {volume} {126}},\
  \bibinfo {pages} {233404} (\bibinfo {year} {2021})}\BibitemShut {NoStop}%
\bibitem [{\citenamefont {Dereli}\ \emph {et~al.}(2012)\citenamefont {Dereli},
  \citenamefont {G\"ul}, \citenamefont {Forn-D\'{\i}az},\ and\ \citenamefont
  {M\"ustecapl\ifmmode \imath \else \i \fi{}o\ifmmode~\breve{g}\else
  \u{g}\fi{}lu}}]{Dereli}%
  \BibitemOpen
  \bibfield  {author} {\bibinfo {author} {\bibfnamefont {T.}~\bibnamefont
  {Dereli}}, \bibinfo {author} {\bibfnamefont {Y.}~\bibnamefont {G\"ul}},
  \bibinfo {author} {\bibfnamefont {P.}~\bibnamefont {Forn-D\'{\i}az}},\ and\
  \bibinfo {author} {\bibfnamefont {O.~E.}\ \bibnamefont {M\"ustecapl\ifmmode
  \imath \else \i \fi{}o\ifmmode~\breve{g}\else \u{g}\fi{}lu}},\ }\bibfield
  {title} {\bibinfo {title} {Two-frequency {J}ahn-{T}eller systems in circuit
  {QED}},\ }\href {https://doi.org/10.1103/PhysRevA.85.053841} {\bibfield
  {journal} {\bibinfo  {journal} {Phys. Rev. A}\ }\textbf {\bibinfo {volume}
  {85}},\ \bibinfo {pages} {053841} (\bibinfo {year} {2012})}\BibitemShut
  {NoStop}%
\bibitem [{\citenamefont {Larson}(2008)}]{Larson}%
  \BibitemOpen
  \bibfield  {author} {\bibinfo {author} {\bibfnamefont {J.}~\bibnamefont
  {Larson}},\ }\bibfield  {title} {\bibinfo {title} {Jahn-{T}eller systems from
  a cavity {QED} perspective},\ }\href
  {https://doi.org/10.1103/PhysRevA.78.033833} {\bibfield  {journal} {\bibinfo
  {journal} {Phys. Rev. A}\ }\textbf {\bibinfo {volume} {78}},\ \bibinfo
  {pages} {033833} (\bibinfo {year} {2008})}\BibitemShut {NoStop}%
\bibitem [{\citenamefont {Wang}\ \emph {et~al.}(2023)\citenamefont {Wang},
  \citenamefont {Frattini}, \citenamefont {Chapman}, \citenamefont {Puri},
  \citenamefont {Girvin}, \citenamefont {Devoret},\ and\ \citenamefont
  {Schoelkopf}}]{Christopher2022}%
  \BibitemOpen
  \bibfield  {author} {\bibinfo {author} {\bibfnamefont {C.~S.}\ \bibnamefont
  {Wang}}, \bibinfo {author} {\bibfnamefont {N.~E.}\ \bibnamefont {Frattini}},
  \bibinfo {author} {\bibfnamefont {B.~J.}\ \bibnamefont {Chapman}}, \bibinfo
  {author} {\bibfnamefont {S.}~\bibnamefont {Puri}}, \bibinfo {author}
  {\bibfnamefont {S.~M.}\ \bibnamefont {Girvin}}, \bibinfo {author}
  {\bibfnamefont {M.~H.}\ \bibnamefont {Devoret}},\ and\ \bibinfo {author}
  {\bibfnamefont {R.~J.}\ \bibnamefont {Schoelkopf}},\ }\bibfield  {title}
  {\bibinfo {title} {Observation of wave-packet branching through an engineered
  conical intersection},\ }\href {https://doi.org/10.1103/PhysRevX.13.011008}
  {\bibfield  {journal} {\bibinfo  {journal} {Phys. Rev. X}\ }\textbf {\bibinfo
  {volume} {13}},\ \bibinfo {pages} {011008} (\bibinfo {year}
  {2023})}\BibitemShut {NoStop}%
\bibitem [{\citenamefont {MacDonell}\ \emph {et~al.}(2021)\citenamefont
  {MacDonell}, \citenamefont {Dickerson}, \citenamefont {Birch}, \citenamefont
  {Kumar}, \citenamefont {Edmunds}, \citenamefont {Biercuk}, \citenamefont
  {Hempel},\ and\ \citenamefont {Kassal}}]{MacDonell2021}%
  \BibitemOpen
  \bibfield  {author} {\bibinfo {author} {\bibfnamefont {R.~J.}\ \bibnamefont
  {MacDonell}}, \bibinfo {author} {\bibfnamefont {C.~E.}\ \bibnamefont
  {Dickerson}}, \bibinfo {author} {\bibfnamefont {C.~J.~T.}\ \bibnamefont
  {Birch}}, \bibinfo {author} {\bibfnamefont {A.}~\bibnamefont {Kumar}},
  \bibinfo {author} {\bibfnamefont {C.~L.}\ \bibnamefont {Edmunds}}, \bibinfo
  {author} {\bibfnamefont {M.~J.}\ \bibnamefont {Biercuk}}, \bibinfo {author}
  {\bibfnamefont {C.}~\bibnamefont {Hempel}},\ and\ \bibinfo {author}
  {\bibfnamefont {I.}~\bibnamefont {Kassal}},\ }\bibfield  {title} {\bibinfo
  {title} {Analog quantum simulation of chemical dynamics},\ }\href
  {https://doi.org/10.1039/d1sc02142g} {\bibfield  {journal} {\bibinfo
  {journal} {Chem. Sci.}\ }\textbf {\bibinfo {volume} {12}},\ \bibinfo {pages}
  {9794} (\bibinfo {year} {2021})}\BibitemShut {NoStop}%
\bibitem [{\citenamefont {MacDonell}\ \emph {et~al.}(2022)\citenamefont
  {MacDonell}, \citenamefont {Navickas}, \citenamefont {Wohlers-Reichel},
  \citenamefont {Valahu}, \citenamefont {Rao}, \citenamefont {Millican},
  \citenamefont {Currington}, \citenamefont {Biercuk}, \citenamefont {Tan},
  \citenamefont {Hempel},\ and\ \citenamefont {Kassal}}]{MacDonell2022}%
  \BibitemOpen
  \bibfield  {author} {\bibinfo {author} {\bibfnamefont {R.~J.}\ \bibnamefont
  {MacDonell}}, \bibinfo {author} {\bibfnamefont {T.}~\bibnamefont {Navickas}},
  \bibinfo {author} {\bibfnamefont {T.~F.}\ \bibnamefont {Wohlers-Reichel}},
  \bibinfo {author} {\bibfnamefont {C.~H.}\ \bibnamefont {Valahu}}, \bibinfo
  {author} {\bibfnamefont {A.~D.}\ \bibnamefont {Rao}}, \bibinfo {author}
  {\bibfnamefont {M.~J.}\ \bibnamefont {Millican}}, \bibinfo {author}
  {\bibfnamefont {M.~A.}\ \bibnamefont {Currington}}, \bibinfo {author}
  {\bibfnamefont {M.~J.}\ \bibnamefont {Biercuk}}, \bibinfo {author}
  {\bibfnamefont {T.~R.}\ \bibnamefont {Tan}}, \bibinfo {author} {\bibfnamefont
  {C.}~\bibnamefont {Hempel}},\ and\ \bibinfo {author} {\bibfnamefont
  {I.}~\bibnamefont {Kassal}},\ }\bibfield  {title} {\bibinfo {title}
  {Predicting molecular vibronic spectra using time-domain analog quantum
  simulation},\ }\Eprint {https://arxiv.org/abs/2209.06558} {arXiv:2209.06558}
  (\bibinfo {year} {2022})\BibitemShut {NoStop}%
\bibitem [{\citenamefont {Bersuker}(2001)}]{Bersuker}%
  \BibitemOpen
  \bibfield  {author} {\bibinfo {author} {\bibfnamefont {I.~B.}\ \bibnamefont
  {Bersuker}},\ }\bibfield  {title} {\bibinfo {title} {Modern aspects of the
  {J}ahn-{T}eller effect: {T}heory and applications to molecular problems},\
  }\href {https://doi.org/10.1021/cr0004411} {\bibfield  {journal} {\bibinfo
  {journal} {Chem. Rev.}\ }\textbf {\bibinfo {volume} {101}},\ \bibinfo {pages}
  {1067} (\bibinfo {year} {2001})}\BibitemShut {NoStop}%
\bibitem [{\citenamefont {Monroe}\ \emph {et~al.}(1996)\citenamefont {Monroe},
  \citenamefont {Meekhof}, \citenamefont {King},\ and\ \citenamefont
  {Wineland}}]{Monroe1996}%
  \BibitemOpen
  \bibfield  {author} {\bibinfo {author} {\bibfnamefont {C.}~\bibnamefont
  {Monroe}}, \bibinfo {author} {\bibfnamefont {D.~M.}\ \bibnamefont {Meekhof}},
  \bibinfo {author} {\bibfnamefont {B.~E.}\ \bibnamefont {King}},\ and\
  \bibinfo {author} {\bibfnamefont {D.~J.}\ \bibnamefont {Wineland}},\
  }\bibfield  {title} {\bibinfo {title} {A ``{S}chr{\"{o}}dinger cat''
  superposition state of an atom},\ }\href
  {https://doi.org/10.1126/science.272.5265.1131} {\bibfield  {journal}
  {\bibinfo  {journal} {Science}\ }\textbf {\bibinfo {volume} {272}},\ \bibinfo
  {pages} {1131} (\bibinfo {year} {1996})}\BibitemShut {NoStop}%
\bibitem [{\citenamefont {Mizrahi}\ \emph {et~al.}(2013)\citenamefont
  {Mizrahi}, \citenamefont {Neyenhuis}, \citenamefont {Johnson}, \citenamefont
  {Campbell}, \citenamefont {Senko}, \citenamefont {Hayes},\ and\ \citenamefont
  {Monroe}}]{Mizrahi2013}%
  \BibitemOpen
  \bibfield  {author} {\bibinfo {author} {\bibfnamefont {J.}~\bibnamefont
  {Mizrahi}}, \bibinfo {author} {\bibfnamefont {B.}~\bibnamefont {Neyenhuis}},
  \bibinfo {author} {\bibfnamefont {K.~G.}\ \bibnamefont {Johnson}}, \bibinfo
  {author} {\bibfnamefont {W.~C.}\ \bibnamefont {Campbell}}, \bibinfo {author}
  {\bibfnamefont {C.}~\bibnamefont {Senko}}, \bibinfo {author} {\bibfnamefont
  {D.}~\bibnamefont {Hayes}},\ and\ \bibinfo {author} {\bibfnamefont
  {C.}~\bibnamefont {Monroe}},\ }\bibfield  {title} {\bibinfo {title} {Quantum
  control of qubits and atomic motion using ultrafast laser pulses},\ }\href
  {https://doi.org/10.1007/s00340-013-5717-6} {\bibfield  {journal} {\bibinfo
  {journal} {Appl. Phys. B}\ }\textbf {\bibinfo {volume} {114}},\ \bibinfo
  {pages} {45} (\bibinfo {year} {2013})}\BibitemShut {NoStop}%
\bibitem [{\citenamefont {Leibfried}\ \emph {et~al.}(1996)\citenamefont
  {Leibfried}, \citenamefont {Meekhof}, \citenamefont {King}, \citenamefont
  {Monroe}, \citenamefont {Itano},\ and\ \citenamefont
  {Wineland}}]{Leibfried1996}%
  \BibitemOpen
  \bibfield  {author} {\bibinfo {author} {\bibfnamefont {D.}~\bibnamefont
  {Leibfried}}, \bibinfo {author} {\bibfnamefont {D.~M.}\ \bibnamefont
  {Meekhof}}, \bibinfo {author} {\bibfnamefont {B.~E.}\ \bibnamefont {King}},
  \bibinfo {author} {\bibfnamefont {C.}~\bibnamefont {Monroe}}, \bibinfo
  {author} {\bibfnamefont {W.~M.}\ \bibnamefont {Itano}},\ and\ \bibinfo
  {author} {\bibfnamefont {D.~J.}\ \bibnamefont {Wineland}},\ }\bibfield
  {title} {\bibinfo {title} {Experimental determination of the motional quantum
  state of a trapped atom},\ }\href
  {https://doi.org/10.1103/physrevlett.77.4281} {\bibfield  {journal} {\bibinfo
   {journal} {Phys. Rev. Lett.}\ }\textbf {\bibinfo {volume} {77}},\ \bibinfo
  {pages} {4281} (\bibinfo {year} {1996})}\BibitemShut {NoStop}%
\bibitem [{\citenamefont {Gerritsma}\ \emph {et~al.}(2010)\citenamefont
  {Gerritsma}, \citenamefont {Kirchmair}, \citenamefont {Z\"{a}hringer},
  \citenamefont {Solano}, \citenamefont {Blatt},\ and\ \citenamefont
  {Roos}}]{Gerritsma2010}%
  \BibitemOpen
  \bibfield  {author} {\bibinfo {author} {\bibfnamefont {R.}~\bibnamefont
  {Gerritsma}}, \bibinfo {author} {\bibfnamefont {G.}~\bibnamefont
  {Kirchmair}}, \bibinfo {author} {\bibfnamefont {F.}~\bibnamefont
  {Z\"{a}hringer}}, \bibinfo {author} {\bibfnamefont {E.}~\bibnamefont
  {Solano}}, \bibinfo {author} {\bibfnamefont {R.}~\bibnamefont {Blatt}},\ and\
  \bibinfo {author} {\bibfnamefont {C.~F.}\ \bibnamefont {Roos}},\ }\bibfield
  {title} {\bibinfo {title} {Quantum simulation of the {D}irac equation},\
  }\href {https://doi.org/10.1038/nature08688} {\bibfield  {journal} {\bibinfo
  {journal} {Nature}\ }\textbf {\bibinfo {volume} {463}},\ \bibinfo {pages}
  {68} (\bibinfo {year} {2010})}\BibitemShut {NoStop}%
\bibitem [{\citenamefont {Johnson}\ \emph {et~al.}(2015)\citenamefont
  {Johnson}, \citenamefont {Neyenhuis}, \citenamefont {Mizrahi}, \citenamefont
  {Wong-Campos},\ and\ \citenamefont {Monroe}}]{Johnson2015}%
  \BibitemOpen
  \bibfield  {author} {\bibinfo {author} {\bibfnamefont {K.~G.}\ \bibnamefont
  {Johnson}}, \bibinfo {author} {\bibfnamefont {B.}~\bibnamefont {Neyenhuis}},
  \bibinfo {author} {\bibfnamefont {J.}~\bibnamefont {Mizrahi}}, \bibinfo
  {author} {\bibfnamefont {J.~D.}\ \bibnamefont {Wong-Campos}},\ and\ \bibinfo
  {author} {\bibfnamefont {C.}~\bibnamefont {Monroe}},\ }\bibfield  {title}
  {\bibinfo {title} {Sensing atomic motion from the zero point to room
  temperature with ultrafast atom interferometry},\ }\href
  {https://doi.org/10.1103/physrevlett.115.213001} {\bibfield  {journal}
  {\bibinfo  {journal} {Phys. Rev. Lett.}\ }\textbf {\bibinfo {volume} {115}},\
  \bibinfo {pages} {213001} (\bibinfo {year} {2015})}\BibitemShut {NoStop}%
\bibitem [{\citenamefont {Fl\"uhmann}\ and\ \citenamefont
  {Home}(2020)}]{Home2020}%
  \BibitemOpen
  \bibfield  {author} {\bibinfo {author} {\bibfnamefont {C.}~\bibnamefont
  {Fl\"uhmann}}\ and\ \bibinfo {author} {\bibfnamefont {J.~P.}\ \bibnamefont
  {Home}},\ }\bibfield  {title} {\bibinfo {title} {Direct
  characteristic-function tomography of quantum states of the trapped-ion
  motional oscillator},\ }\href
  {https://doi.org/10.1103/PhysRevLett.125.043602} {\bibfield  {journal}
  {\bibinfo  {journal} {Phys. Rev. Lett.}\ }\textbf {\bibinfo {volume} {125}},\
  \bibinfo {pages} {043602} (\bibinfo {year} {2020})}\BibitemShut {NoStop}%
\bibitem [{\citenamefont {Jia}\ \emph {et~al.}(2022)\citenamefont {Jia},
  \citenamefont {Wang}, \citenamefont {Zhang}, \citenamefont {Whitlow},
  \citenamefont {Fang}, \citenamefont {Kim},\ and\ \citenamefont
  {Brown}}]{Jia2022}%
  \BibitemOpen
  \bibfield  {author} {\bibinfo {author} {\bibfnamefont {Z.}~\bibnamefont
  {Jia}}, \bibinfo {author} {\bibfnamefont {Y.}~\bibnamefont {Wang}}, \bibinfo
  {author} {\bibfnamefont {B.}~\bibnamefont {Zhang}}, \bibinfo {author}
  {\bibfnamefont {J.}~\bibnamefont {Whitlow}}, \bibinfo {author} {\bibfnamefont
  {C.}~\bibnamefont {Fang}}, \bibinfo {author} {\bibfnamefont {J.}~\bibnamefont
  {Kim}},\ and\ \bibinfo {author} {\bibfnamefont {K.~R.}\ \bibnamefont
  {Brown}},\ }\bibfield  {title} {\bibinfo {title} {Determination of multimode
  motional quantum states in a trapped ion system},\ }\href
  {https://doi.org/10.1103/PhysRevLett.129.103602} {\bibfield  {journal}
  {\bibinfo  {journal} {Phys. Rev. Lett.}\ }\textbf {\bibinfo {volume} {129}},\
  \bibinfo {pages} {103602} (\bibinfo {year} {2022})}\BibitemShut {NoStop}%
\bibitem [{\citenamefont {Hayes}\ \emph {et~al.}(2014)\citenamefont {Hayes},
  \citenamefont {Flammia},\ and\ \citenamefont {Biercuk}}]{Hayes_2014}%
  \BibitemOpen
  \bibfield  {author} {\bibinfo {author} {\bibfnamefont {D.}~\bibnamefont
  {Hayes}}, \bibinfo {author} {\bibfnamefont {S.~T.}\ \bibnamefont {Flammia}},\
  and\ \bibinfo {author} {\bibfnamefont {M.~J.}\ \bibnamefont {Biercuk}},\
  }\bibfield  {title} {\bibinfo {title} {Programmable quantum simulation by
  dynamic {H}amiltonian engineering},\ }\href
  {https://doi.org/10.1088/1367-2630/16/8/083027} {\bibfield  {journal}
  {\bibinfo  {journal} {New J. Phys.}\ }\textbf {\bibinfo {volume} {16}},\
  \bibinfo {pages} {083027} (\bibinfo {year} {2014})}\BibitemShut {NoStop}%
\bibitem [{\citenamefont {Brownnutt}\ \emph {et~al.}(2015)\citenamefont
  {Brownnutt}, \citenamefont {Kumph}, \citenamefont {Rabl},\ and\ \citenamefont
  {Blatt}}]{Brownnutt2015}%
  \BibitemOpen
  \bibfield  {author} {\bibinfo {author} {\bibfnamefont {M.}~\bibnamefont
  {Brownnutt}}, \bibinfo {author} {\bibfnamefont {M.}~\bibnamefont {Kumph}},
  \bibinfo {author} {\bibfnamefont {P.}~\bibnamefont {Rabl}},\ and\ \bibinfo
  {author} {\bibfnamefont {R.}~\bibnamefont {Blatt}},\ }\bibfield  {title}
  {\bibinfo {title} {Ion-trap measurements of electric-field noise near
  surfaces},\ }\href {https://doi.org/10.1103/revmodphys.87.1419} {\bibfield
  {journal} {\bibinfo  {journal} {Rev. Mod. Phys.}\ }\textbf {\bibinfo {volume}
  {87}},\ \bibinfo {pages} {1419} (\bibinfo {year} {2015})}\BibitemShut
  {NoStop}%
\bibitem [{\citenamefont {Kienzler}\ \emph {et~al.}(2016)\citenamefont
  {Kienzler}, \citenamefont {Fl\"uhmann}, \citenamefont {Negnevitsky},
  \citenamefont {Lo}, \citenamefont {Marinelli}, \citenamefont {Nadlinger},\
  and\ \citenamefont {Home}}]{Kienzler2016}%
  \BibitemOpen
  \bibfield  {author} {\bibinfo {author} {\bibfnamefont {D.}~\bibnamefont
  {Kienzler}}, \bibinfo {author} {\bibfnamefont {C.}~\bibnamefont
  {Fl\"uhmann}}, \bibinfo {author} {\bibfnamefont {V.}~\bibnamefont
  {Negnevitsky}}, \bibinfo {author} {\bibfnamefont {H.-Y.}\ \bibnamefont {Lo}},
  \bibinfo {author} {\bibfnamefont {M.}~\bibnamefont {Marinelli}}, \bibinfo
  {author} {\bibfnamefont {D.}~\bibnamefont {Nadlinger}},\ and\ \bibinfo
  {author} {\bibfnamefont {J.~P.}\ \bibnamefont {Home}},\ }\bibfield  {title}
  {\bibinfo {title} {Observation of quantum interference between separated
  mechanical oscillator wave packets},\ }\href
  {https://doi.org/10.1103/PhysRevLett.116.140402} {\bibfield  {journal}
  {\bibinfo  {journal} {Phys. Rev. Lett.}\ }\textbf {\bibinfo {volume} {116}},\
  \bibinfo {pages} {140402} (\bibinfo {year} {2016})}\BibitemShut {NoStop}%
\bibitem [{\citenamefont {Whitlow}\ \emph {et~al.}(2022)\citenamefont
  {Whitlow}, \citenamefont {Jia}, \citenamefont {Wang}, \citenamefont {Fang},
  \citenamefont {Kim},\ and\ \citenamefont {Brown}}]{Whitlow2022}%
  \BibitemOpen
  \bibfield  {author} {\bibinfo {author} {\bibfnamefont {J.}~\bibnamefont
  {Whitlow}}, \bibinfo {author} {\bibfnamefont {Z.}~\bibnamefont {Jia}},
  \bibinfo {author} {\bibfnamefont {Y.}~\bibnamefont {Wang}}, \bibinfo {author}
  {\bibfnamefont {C.}~\bibnamefont {Fang}}, \bibinfo {author} {\bibfnamefont
  {J.}~\bibnamefont {Kim}},\ and\ \bibinfo {author} {\bibfnamefont {K.~R.}\
  \bibnamefont {Brown}},\ }\bibfield  {title} {\bibinfo {title} {Simulating
  conical intersections with trapped ions},\ }\Eprint
  {https://arxiv.org/abs/2211.07319} {arXiv:2211.07319}  (\bibinfo {year}
  {2022})\BibitemShut {NoStop}%
\end{thebibliography}

\begin{thebibliography}{55}%
\makeatletter
\providecommand \@ifxundefined [1]{%
 \@ifx{#1\undefined}
}%
\providecommand \@ifnum [1]{%
 \ifnum #1\expandafter \@firstoftwo
 \else \expandafter \@secondoftwo
 \fi
}%
\providecommand \@ifx [1]{%
 \ifx #1\expandafter \@firstoftwo
 \else \expandafter \@secondoftwo
 \fi
}%
\providecommand \natexlab [1]{#1}%
\providecommand \enquote  [1]{``#1''}%
\providecommand \bibnamefont  [1]{#1}%
\providecommand \bibfnamefont [1]{#1}%
\providecommand \citenamefont [1]{#1}%
\providecommand \href@noop [0]{\@secondoftwo}%
\providecommand \href [0]{\begingroup \@sanitize@url \@href}%
\providecommand \@href[1]{\@@startlink{#1}\@@href}%
\providecommand \@@href[1]{\endgroup#1\@@endlink}%
\providecommand \@sanitize@url [0]{\catcode `\\12\catcode `\$12\catcode
  `\&12\catcode `\#12\catcode `\^12\catcode `\_12\catcode `\%12\relax}%
\providecommand \@@startlink[1]{}%
\providecommand \@@endlink[0]{}%
\providecommand \url  [0]{\begingroup\@sanitize@url \@url }%
\providecommand \@url [1]{\endgroup\@href {#1}{\urlprefix }}%
\providecommand \urlprefix  [0]{URL }%
\providecommand \Eprint [0]{\href }%
\providecommand \doibase [0]{https://doi.org/}%
\providecommand \selectlanguage [0]{\@gobble}%
\providecommand \bibinfo  [0]{\@secondoftwo}%
\providecommand \bibfield  [0]{\@secondoftwo}%
\providecommand \translation [1]{[#1]}%
\providecommand \BibitemOpen [0]{}%
\providecommand \bibitemStop [0]{}%
\providecommand \bibitemNoStop [0]{.\EOS\space}%
\providecommand \EOS [0]{\spacefactor3000\relax}%
\providecommand \BibitemShut  [1]{\csname bibitem#1\endcsname}%
\let\auto@bib@innerbib\@empty
\addtocounter{NAT@ctr}{51}
%</preamble>
\bibitem [{\citenamefont {Monroe}\ \emph {et~al.}(1995)\citenamefont {Monroe},
  \citenamefont {Meekhof}, \citenamefont {King}, \citenamefont {Jefferts},
  \citenamefont {Itano}, \citenamefont {Wineland},\ and\ \citenamefont
  {Gould}}]{Monroe1995}%
  \BibitemOpen
  \bibfield  {author} {\bibinfo {author} {\bibfnamefont {C.}~\bibnamefont
  {Monroe}}, \bibinfo {author} {\bibfnamefont {D.~M.}\ \bibnamefont {Meekhof}},
  \bibinfo {author} {\bibfnamefont {B.~E.}\ \bibnamefont {King}}, \bibinfo
  {author} {\bibfnamefont {S.~R.}\ \bibnamefont {Jefferts}}, \bibinfo {author}
  {\bibfnamefont {W.~M.}\ \bibnamefont {Itano}}, \bibinfo {author}
  {\bibfnamefont {D.~J.}\ \bibnamefont {Wineland}},\ and\ \bibinfo {author}
  {\bibfnamefont {P.}~\bibnamefont {Gould}},\ }\bibfield  {title} {\bibinfo
  {title} {Resolved-sideband {R}aman cooling of a bound atom to the 3{D}
  zero-point energy},\ }\href {https://doi.org/10.1103/physrevlett.75.4011}
  {\bibfield  {journal} {\bibinfo  {journal} {Phys. Rev. Lett.}\ }\textbf
  {\bibinfo {volume} {75}},\ \bibinfo {pages} {4011} (\bibinfo {year}
  {1995})}\BibitemShut {NoStop}%
\bibitem [{\citenamefont {Wineland}\ \emph {et~al.}(1998)\citenamefont
  {Wineland}, \citenamefont {Monroe}, \citenamefont {Itano}, \citenamefont
  {Leibfried}, \citenamefont {King},\ and\ \citenamefont
  {Meekhof}}]{Wineland1998}%
  \BibitemOpen
  \bibfield  {author} {\bibinfo {author} {\bibfnamefont {D.~J.}\ \bibnamefont
  {Wineland}}, \bibinfo {author} {\bibfnamefont {C.}~\bibnamefont {Monroe}},
  \bibinfo {author} {\bibfnamefont {W.~M.}\ \bibnamefont {Itano}}, \bibinfo
  {author} {\bibfnamefont {D.}~\bibnamefont {Leibfried}}, \bibinfo {author}
  {\bibfnamefont {B.~E.}\ \bibnamefont {King}},\ and\ \bibinfo {author}
  {\bibfnamefont {D.~M.}\ \bibnamefont {Meekhof}},\ }\bibfield  {title}
  {\bibinfo {title} {Experimental issues in coherent quantum-state manipulation
  of trapped atomic ions},\ }\href {https://doi.org/10.6028/jres.103.019}
  {\bibfield  {journal} {\bibinfo  {journal} {J. Res. Natl. Inst. Stand.
  Technol.}\ }\textbf {\bibinfo {volume} {103}},\ \bibinfo {pages} {259}
  (\bibinfo {year} {1998})}\BibitemShut {NoStop}%
\bibitem [{\citenamefont {Riesebos}\ \emph {et~al.}(2021)\citenamefont
  {Riesebos}, \citenamefont {Bondurant},\ and\ \citenamefont
  {Brown}}]{Riesebos2021}%
  \BibitemOpen
  \bibfield  {author} {\bibinfo {author} {\bibfnamefont {L.}~\bibnamefont
  {Riesebos}}, \bibinfo {author} {\bibfnamefont {B.}~\bibnamefont
  {Bondurant}},\ and\ \bibinfo {author} {\bibfnamefont {K.~R.}\ \bibnamefont
  {Brown}},\ }\bibfield  {title} {\bibinfo {title} {Universal graph-based
  scheduling for quantum systems},\ }\href
  {https://doi.org/10.1109/mm.2021.3094968} {\bibfield  {journal} {\bibinfo
  {journal} {{IEEE} Micro}\ }\textbf {\bibinfo {volume} {41}},\ \bibinfo
  {pages} {5} (\bibinfo {year} {2021})}\BibitemShut {NoStop}%
\bibitem [{\citenamefont {Valahu}\ \emph {et~al.}(2023)\citenamefont {Valahu},
  \citenamefont {Olaya-Agudelo}, \citenamefont {MacDonell}, \citenamefont
  {Navickas}, \citenamefont {Rao}, \citenamefont {Millican}, \citenamefont
  {P\'erez-S\'anchez}, \citenamefont {Yuen-Zhou}, \citenamefont {Biercuk},
  \citenamefont {Hempel}, \citenamefont {Tan},\ and\ \citenamefont
  {Kassal}}]{zenodo}%
  \BibitemOpen
  \bibfield  {author} {\bibinfo {author} {\bibfnamefont {C.~H.}\ \bibnamefont
  {Valahu}}, \bibinfo {author} {\bibfnamefont {V.~C.}\ \bibnamefont
  {Olaya-Agudelo}}, \bibinfo {author} {\bibfnamefont {R.~J.}\ \bibnamefont
  {MacDonell}}, \bibinfo {author} {\bibfnamefont {T.}~\bibnamefont {Navickas}},
  \bibinfo {author} {\bibfnamefont {A.~D.}\ \bibnamefont {Rao}}, \bibinfo
  {author} {\bibfnamefont {M.~J.}\ \bibnamefont {Millican}}, \bibinfo {author}
  {\bibfnamefont {J.}~\bibnamefont {P\'erez-S\'anchez}}, \bibinfo {author}
  {\bibfnamefont {J.}~\bibnamefont {Yuen-Zhou}}, \bibinfo {author}
  {\bibfnamefont {M.~J.}\ \bibnamefont {Biercuk}}, \bibinfo {author}
  {\bibfnamefont {C.}~\bibnamefont {Hempel}}, \bibinfo {author} {\bibfnamefont
  {T.~R.}\ \bibnamefont {Tan}},\ and\ \bibinfo {author} {\bibfnamefont
  {I.}~\bibnamefont {Kassal}},\ }\bibfield  {title} {\bibinfo {title} {Direct
  observation of geometric phase in dynamics around a conical intersection [Dataset]},\
  }\bibfield  {journal} {\bibinfo  {journal} {Zenodo}\ }\href
  {https://doi.org/10.5281/zenodo.7955887} {https://doi.org/10.5281/zenodo.7955887} (\bibinfo
  {year} {2023})\BibitemShut {NoStop}%
\end{thebibliography}
\end{document}